%% file: paper_revision_nobold.tex
\documentclass[apjl]{emulateapj}
\usepackage{comment}
\usepackage{ifthen}
\usepackage{lineno}
\usepackage{placeins}
\usepackage{amsmath}

\newcommand{\ctbd}[1]{}

\newcommand{\Lc}{Light curve}

\newcommand{\kms}{\ensuremath{\rm km\,s^{-1}}}

\newcommand{\logg}{\ensuremath{\log{g}}}
\newcommand{\vsini}{\ensuremath{v \sin{I_\star}}}

\newcommand{\thisstar}{MWC 882}

\newcommand{\Mone}{\ensuremath{3.24 \pm 0.29}}
\newcommand{\Mtwo}{\ensuremath{0.542 \pm 0.053}}
\newcommand{\Kone}{\ensuremath{14.3 _{-1.3}^{+1.4}}}
\newcommand{\Ktwo}{\ensuremath{85.5 _{-2.2}^{+1.9}}}
\newcommand{\Rone}{\ensuremath{3.09 \pm 0.59}}
\newcommand{\Rtwo}{\ensuremath{2.01 \pm 0.52}}
\newcommand{\Rdisk}{\ensuremath{59.9 \pm 6.2}}
\newcommand{\Rsep}{\ensuremath{114.0 \pm 3.1}}

\usepackage{color}

\usepackage{ulem}

\newboolean{emulateapj}
\setboolean{emulateapj}{true}

\newboolean{rvtablelong}
\setboolean{rvtablelong}{true}

\newboolean{astroph}
\setboolean{astroph}{true}

\shortauthors{Zhou et al. }
\shorttitle{Disk occultations in a post-Algol system}
\ifthenelse{\boolean{emulateapj}}{
    \newcommand{\titledag}{$\dagger$}
}{
    \newcommand{\titledag}{\dagger}
}
\ifthenelse{\boolean{emulateapj}}{
    \newcommand{\titlestar}{$\star$}
}{
    \newcommand{\titlestar}{\star}
}
\ifthenelse{\boolean{emulateapj}}{
    
}{
    
}

\begin{document}


\title{Occultations from an active accretion disk in a 72\,day detached post-Algol system detected by {\it K2}}

\author{
G. Zhou \altaffilmark{1,\titledag},
S. Rappaport \altaffilmark{2},
L. Nelson \altaffilmark{3},
C.X. Huang \altaffilmark{2},
A. Senhadji \altaffilmark{3},
J.E. Rodriguez \altaffilmark{1},
A. Vanderburg \altaffilmark{1,4,\titlestar},
S. Quinn \altaffilmark{1},
C.I. Johnson \altaffilmark{1},
D.W. Latham \altaffilmark{1},
G. Torres  \altaffilmark{1},
B.L. Gary \altaffilmark{5},
T.G. Tan \altaffilmark{6},
M.C. Johnson \altaffilmark{7},
J. Burt \altaffilmark{2},
M.H. Kristiansen \altaffilmark{8,9},
T.L. Jacobs \altaffilmark{10},
D. LaCourse \altaffilmark{11},
H. M. Schwengeler \altaffilmark{12},
I. Terentev \altaffilmark{13},
A. Bieryla \altaffilmark{1},
G.A. Esquerdo \altaffilmark{1},
P. Berlind \altaffilmark{1},
M.L. Calkins \altaffilmark{1},
J. Bento \altaffilmark{14},
W.D. Cochran \altaffilmark{4},
M. Karjalainen \altaffilmark{15},
A.P. Hatzes \altaffilmark{16},
R. Karjalainen \altaffilmark{15},
B. Holden \altaffilmark{17},
R.P. Butler \altaffilmark{18}
}

\altaffiltext{1}{Harvard-Smithsonian Center for Astrophysics, 60 Garden Street, Cambridge, MA 02138 USA; george.zhou@cfa.harvard.edu}
\altaffiltext{2}{Department of Physics, and Kavli Institute for Astrophysics and Space Research, M.I.T., Cambridge, MA 02139, USA}
\altaffiltext{3}{Department of Physics and Astronomy, Bishop?s University, 2600 College St., Sherbrooke, QC J1M 1Z7}
\altaffiltext{4}{Department of Astronomy, The University of Texas at Austin, 2515 Speedway, Stop C1400, Austin, TX 78712}
\altaffiltext{5}{Hereford Arizona Observatory, Hereford, AZ 85615}
\altaffiltext{6}{Perth Exoplanet Survey Telescope, Perth, Australia}
\altaffiltext{7}{Department of Astronomy, The Ohio State University, Columbus, OH 43210, USA}
\altaffiltext{8}{DTU Space, National Space Institute, Technical University of Denmark, Elektrovej 327, DK-2800 Lyngby, Denmark} 
\altaffiltext{9}{Brorfelde Observatory, Observator Gyldenkernes Vej 7, DK-4340 {T{\o}ll{\o}se}, Denmark} 
\altaffiltext{10}{Amateur Astronomer, 12812 SE 69th Place Bellevue, WA 98006 USA} 
\altaffiltext{11}{Amateur Astronomer, 7507 52nd Place NE Marysville, WA 98270 USA} 
\altaffiltext{12}{Citizen Scientist at planethunters.org, Zehntenfreistrasse 11, CH-4103 Bottmingen, Switzerland} 
\altaffiltext{13}{Citizen Scientist at planethunters.org, Moskovskaya 8, 185031 Petrozavodsk, Russia } 
\altaffiltext{14}{Research School of Astronomy and Astrophysics, Australian National University, Cotter Rd, Weston Creek, ACT, Australia, 2611}
\altaffiltext{15}{Isaac Newton Group of Telescopes, Apartado de Correos 321, Santa Cruz de La Palma, E-38700, Spain}
\altaffiltext{16}{Thuringer Landessternwarte Tautenburg, Sternwarte 5, Tautenburg, D-07778, Germany}
\altaffiltext{17}{UCO/Lick Observatory, Department of Astronomy and Astrophysics, University of California at Santa Cruz, Santa Cruz, CA 95064, USA}
\altaffiltext{18}{Department of Terrestrial Magnetism, Carnegie Institution for Science, Washington, DC 20015, USA}

\altaffiltext{\titledag}{Hubble Fellow}
\altaffiltext{\titlestar}{Sagan Fellow}


\begin{abstract}

Disks in binary systems can cause exotic eclipsing events. \thisstar{} (BD-22 4376, EPIC 225300403) is such a disk-eclipsing system identified from observations during Campaign 11 of the {\it K2} mission. We propose that \thisstar{} is a post-Algol system with a B7 donor star of mass $\Mtwo\,M_\odot$ in a 72\,day period orbit around an A0 accreting star of mass $\Mone\,M_\odot$. The $\Rdisk\,R_\odot$ disk around the accreting star occults the donor star once every orbit, inducing 19\,day long, 7\% deep eclipses identified by {\it K2}, and subsequently found in pre-discovery ASAS and ASAS-SN observations. We coordinated a campaign of photometric and spectroscopic observations for \thisstar{} to measure the dynamical masses of the components and to monitor the system during eclipse. We found the photometric eclipse to be gray to $\approx 1$\%. We found the primary star exhibits spectroscopic signatures of active accretion, and observed gas absorption features from the disk during eclipse. We suggest \thisstar{} initially consisted of a $\approx 3.6\,M_\odot$ donor star transferring mass via Roche lobe overflow to a $\approx 2.1\,M_\odot$ accretor in a $\approx 7$\,day initial orbit. Through angular momentum conservation, the donor star is pushed outward during mass transfer to its current orbit of 72\,days. The observed state of the system corresponds with the donor star having left the Red Giant Branch $\sim 0.3$\,Myr ago, terminating active mass transfer. The present disk is expected to be short-lived ($10^2$ years) without an active feeding mechanism, presenting a challenge to this model.


\setcounter{footnote}{0}
\end{abstract}

\keywords{
    accretion, accretion disks, binaries: eclipsing, stars: evolution
}


\section{Introduction}
\label{sec:introduction}

Approximately 70\% of intermediate mass stars reside in multi-stellar systems \citep[e.g.][]{2013ARA&A..51..269D,2017ApJS..230...15M}, and perhaps only a half of them can go through life without experiencing the influence of their partners. Binary stars are important astrophysical laboratories, where the primordial stars share a common age, and these are excellent test-beds of binary stellar evolution, mass-transfer, and accretion processes. The final states of many binaries are shaped by mass transfer events during their lifetimes. 

Algols are binary stars that experience mass transfer when the primary star in the primordial binary evolves. In these binaries \citep{1989SSRv...50....1B,1989SSRv...50.....B,2001ASSL..264...79P}, the more massive companion evolves off the main sequence first, and expands to fill its Roche lobe. Overflowing matter from the more massive donor star accretes onto the mass-gaining star (hereafter `accretor'), leading to an inverted mass ratio for the resultant system. The actual pathway for each Algol depends on the initial mass ratio of the binary, the initial orbital period, how conservative the mass transfer is, and how much specific angular momentum is carried away with the lost mater \citep[e.g.][]{2001ApJ...552..664N,2002ApJ...575..461E,2005Ap&SS.296..353D}. Some Algols are stable mass-transferring systems and leave remnant thermally bloated white dwarfs \citep[e.g.][]{2015ApJ...803...82R}  or subdwarf B and O stars \citep[e.g.][]{2002MNRAS.336..449H}.  However, shortly after the mass-transfer phase ends, the tenuous residual atmosphere of the mass losing giant can remain bloated for up to a Myr until the remaining hydrogen is consumed, while the envelope shrinks and gets hotter, to the point where the underlying white dwarf or subdwarf is revealed.  This might be called a `transitional phase'.

It may be possible for the accretion disks to persist in systems that have ended active mass transfer. Occultations induced by persistent disks have been detected in a handful of long period binaries. Photometric occultations by disks are often distinct from other events, as they exhibit long duration, deep eclipse-like signals that can be identified in photometric surveys. $\epsilon$ Aurigae is the classical example of such a system, with a post-AGB F0 supergiant with a B-star companion embedded in a dusty disk with an orbital period of 27 years \citep[e.g.][]{1937ApJ....86..570K,1965ApJ...141..976H,1971Ap&SS..10..332K}. The occultation is observed in photometry \citep[e.g.][]{1970VA.....12..199G,1986ApJ...300L..11K,1991ApJ...367..278C}, interferometry \citep{2010Natur.464..870K,2015ApJS..220...14K}, and spectroscopy \citep[e.g.][]{1986PASP...98..389L,2011A&A...530A.146C,2012JAVSO..40..729L,2013PASP..125..775G,2014MNRAS.445.2884M,2014AN....335..904S}, and the system's orbit and masses are constrained by the radial velocities \citep{2010AJ....139.1254S}. The two year long eclipse is inferred to be from a $\sim 4$\,AU diameter disk consisting of both dust and gas \citep[e.g.][]{1996ApJ...465..371L}. The dusty component is required to explain the strong IR excess in the system \citep{2010ApJ...714..549H}, while the gaseous component in the vertically extended disk shell induces a series of spectroscopic features that map out the Keplerian disk during occultation \citep[e.g.][]{2011A&A...530A.146C,2014AN....335..904S}. A central brightening is seen during eclipse, leading to a flared disk geometry interpretation \citep[e.g.][]{2011A&A...532L..12B}. Similar long period, long duration disk occultations in other potential mass-transfer systems have also been identified: the 96 day period V383 Sco \citep{1994A&A...285..531Z}, the 5.6 year period EE Cep \citep{1999MNRAS.303..521M}, the 468 day periodiocally recurring eclipses of OGLE-LMC-ECL-11893 \citep{2014ApJ...788...41D,2014ApJ...797....6S}, and 1277 day periodic eclipses of OGLE-BLG182.1.162852 \citep{2015MNRAS.447L..31R}. The recent advent of wide field photometric surveys is now also enabling numerous new discoveries, including the 69 year eclipsing disk system TYC 2505-672-1 \citep{2016AJ....151..123R}. 

\begin{figure}[!ht]
\includegraphics[width=\linewidth]{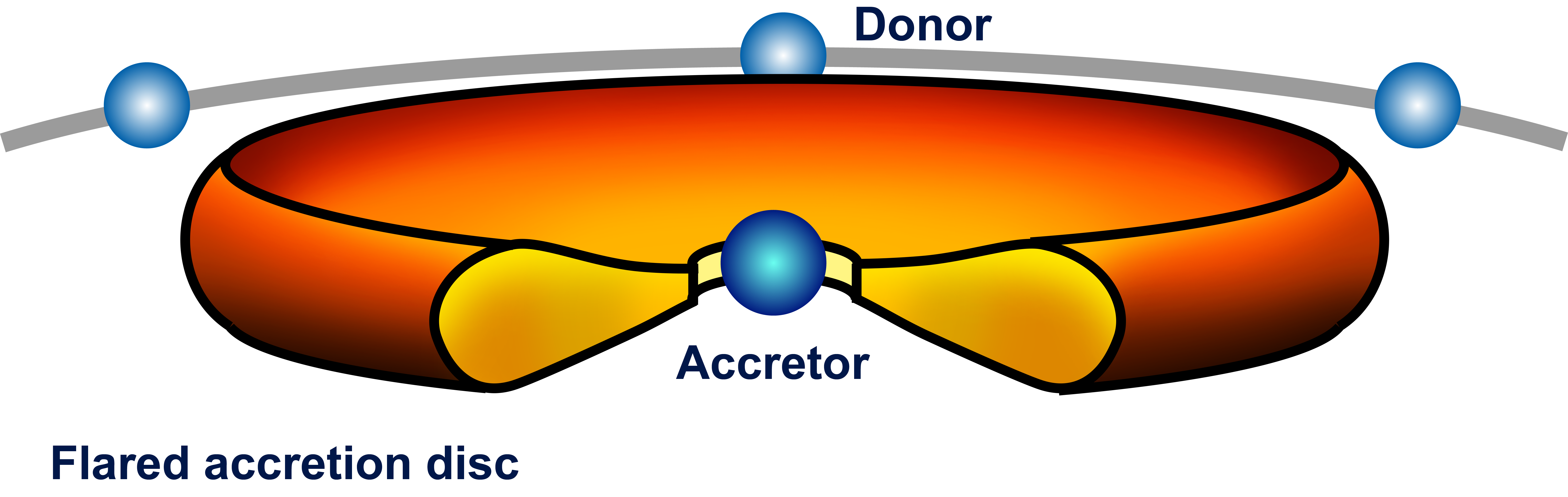}
\caption{
In the \thisstar{} system, the $3.2\,M_\odot$ A0 accreting star is enshrouded in an accretion disk $\approx 60\,R_\odot$ in radius, orbited by a post-RGB $0.5\,M_\odot$ B7 donor star with an orbital period of 72\,days at a separation of $114\, R_\odot$. The donor star is occulted by the accretion disk every orbit, causing periodic eclipses $\approx 7$\% deep. Our observations suggest that the donor star is much smaller than its Roche lobe, and is not sustaining active mass transfer. Illustration shows the geometry of the system from the observer, and is not to scale.
\label{fig:disk_illustration}} 
\end{figure}

\thisstar{} ($V_\mathrm{mag} = 10.8$) was originally identified in the Mount Wilson Catalog (MWC) of A and B stars due to its Balmer line emissions \citep{1949ApJ...110..387M}.  Here we report that \thisstar{} is a 72-day period binary involving a B7 post Red Giant Branch (RGB) star and an A0 accretor, exhibiting disk occultations on each orbit. Occultations of the donor by the accretion disk around the accretor were detected by the {\it K2} mission (during Campaign 11), and subsequently identified in pre-discovery observations by ground-based photometric surveys. The basic geometry of the system is illustrated in Figure~\ref{fig:disk_illustration}. As we detail below, the disk signatures of \thisstar{} are eerily reminiscent of $\epsilon$ Aurigae. The eclipse light curve exhibits a central brightening that we also argue to be from a flared-disk geometry. Spectroscopic signatures of the disk were detected, via a similar set of absorption lines, as that from the recent $\epsilon$ Aurigae eclipse. Unlike $\epsilon$ Aurigae, both members of the system are directly detected, and their dynamical masses can be measured. Our interpretation of \thisstar{}, as a post-Algol with a donor star that relatively recently ended its mass transfer phase, will aid the understanding of similar long period disk occultation systems.

\section{Detection and Photometric follow-up observations}
\label{sec:Photobs}

\begin{figure*}[!ht]
\plotone{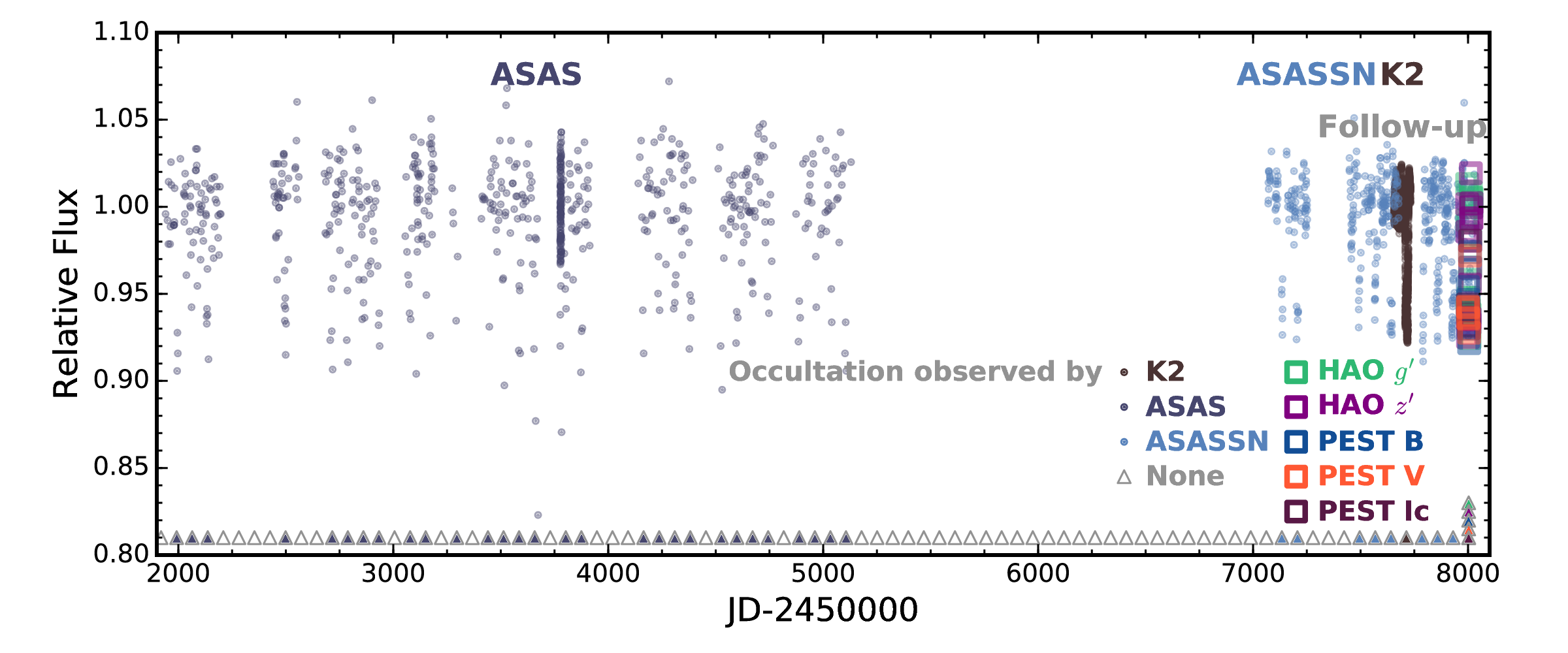}\\
\plotone{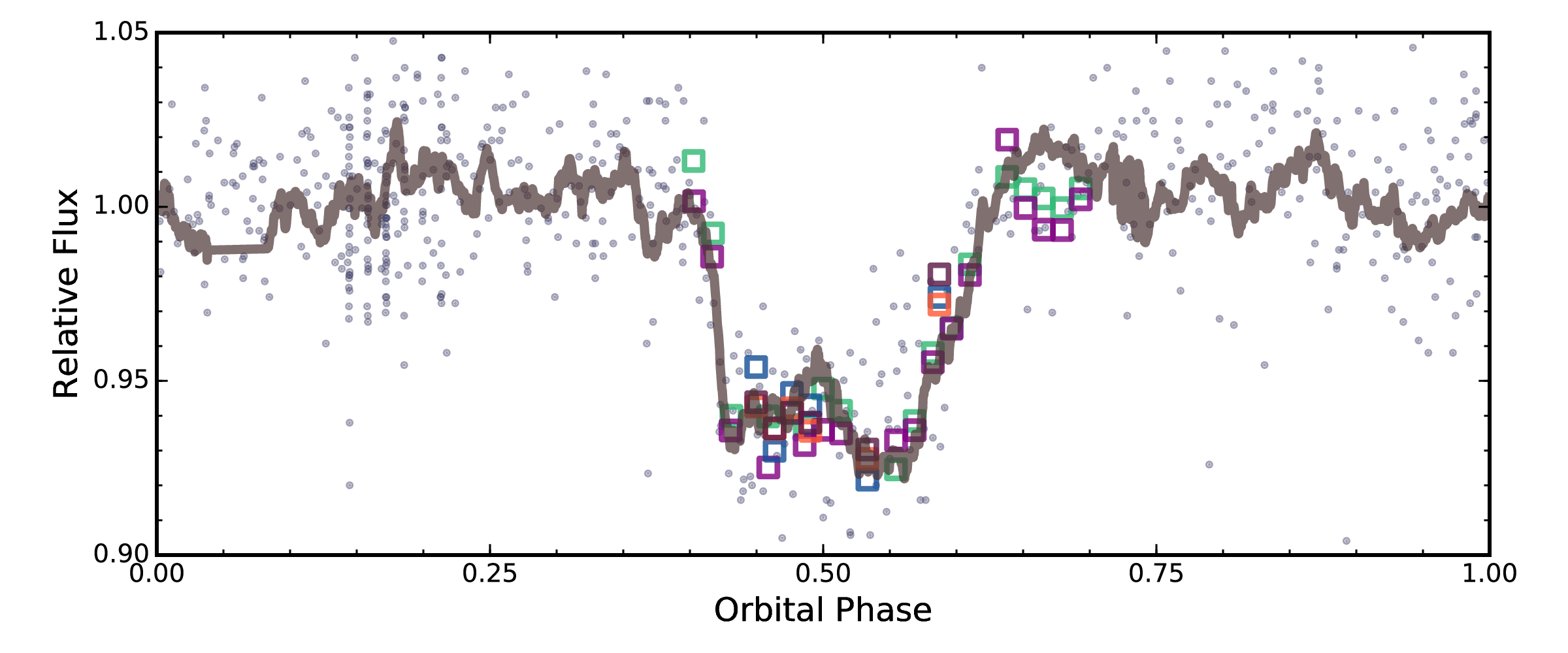}

\caption{
The collection of photometry available for \thisstar{}. \textbf{Top} panel shows the observations spanning 16 years, from ASAS and ASAS-SN pre-discovery survey photometry to follow-up observations by PEST and HAO. Individual occultation epochs are labelled by the triangles at the bottom of the panel, color-coded by facility. Epochs that were not observed are labelled by open triangles. \textbf{Bottom} panel shows the same set of photometry, phase folded such that the occultation is at phase 0.5. The {\it K2} light curve is marked by the brown line, multi-band follow-up photometry by open squares (binned to the per night median), and pre-covery observations from ASAS and ASAS-SN by the solid circles. Note the lack of color dependence in the multi-band follow-up observations, to within $\sim 1$\% precision. 
\label{fig:lc}} 
\end{figure*}

The photometric and astrometric properties of \thisstar{} are listed in Table~\ref{tab:litparams}. A total of 38 photometric occultations were recorded in multiple datasets, including the {\it K2} discovery light curves, pre-discovery datasets from ground-based wide-field photometric surveys, and our multi-wavelength follow-up observations. They are summarized below, listed in Table~\ref{tab:photobs} for clarity, and plotted in Figure~\ref{fig:lc} in full.

\input{literature_params.tex}

\begin{deluxetable*}{llrr}
\tablewidth{0pc}
\tabletypesize{\scriptsize}
\tablecaption{
    Summary of photometric observations
    \label{tab:photobs}
}
\tablehead{
    \multicolumn{1}{c}{Facility}          &
    \multicolumn{1}{c}{Date(s)}             &
    \multicolumn{1}{c}{Number of Images}     &
    \multicolumn{1}{c}{Filter}            
}
\startdata
ASAS & 2001 Jan 31 -- 2009 Oct 25 & 662 & $V$\\
ASAS-SN & 2015 Feb 16 -- 2017 Jul 2 & 418 & $V$\\
{\it K2} & 2016 Sep 24 -- 2016 Oct 18 & 1062 & $Kp$ \\
{\it K2} & 2016 Oct 21 -- 2016 Dec 07 & 2174 & $Kp$ \\
HAO & 2017 Aug 30 -- 2017 Sep 20 & 1000 & $g'$ \\
HAO & 2017 Aug 30 -- 2017 Sep 20 & 870 & $z'$ \\
PEST & 2017 Sep 2 -- 2017 Sep 12 & 79 & $B$ \\
PEST & 2017 Sep 2 -- 2017 Sep 12 & 79 & $V$ \\
PEST & 2017 Sep 2 -- 2017 Sep 12 & 83 & $I$ \\
\enddata 

\end{deluxetable*}

\subsection{Identification from {\it K2} photometry}
\label{sec:K2}

\thisstar{} was observed during Campaign 11 of the {\it K2} mission with the {\it Kepler} spacecraft \citep{2014PASP..126..398H} under the designation EPIC 225300403. {\it K2} provides photometric coverage of selected stars in fields distributed across the ecliptic plane over a temporal baseline of $\approx 72$ days per field. \thisstar{} was observed with a cadence of $30$\,min by {\it K2}. The target pixels were downloaded upon public release of the campaign data, and reduced as per \citet{2014PASP..126..948V}\footnote{After identifying \thisstar\ as a target of interest, we re-reduced the light curve allowing low-frequency variations to be modeled with a faster basis spline, which yielded a light curve with fewer systematics.}.

We conducted a visual examination of Campaign 11 light curves for astrophysical features not usually identified by automated algorithms and periodic signal analyses. The inspections are aided by the {\it LCTOOLS} software and {\it LCViewer} packages, and was conducted in the fashion described in \citet{2017arXiv170806069R}. The light curve of \thisstar{} revealed a transit-like feature, with a depth of $\approx 7$\%, and a full duration of $\approx 19.4$ days (see Figure~\ref{fig:lc}). The occultation is slightly asymmetric, with `ingress' spanning over 2 days, and the egress extending over 8 days. A $\approx 3$\% brightening is seen at the center of the eclipse.

\subsection{Pre-covery with ASAS and ASAS-SN}
\label{sec:precovery}

Once the eclipse of \thisstar{} was identified in the {\it K2} data, we searched for additional events in archival photometric observations from the ASAS and ASAS-SN surveys. Here we present a brief overview of each survey and the observations available on \thisstar. 

Designed to survey the entire sky and catalog all variable stars brighter than $V_\mathrm{mag}=14$, the All-Sky Automated Survey \citep[ASAS,][]{1997AcA....47..467P,2001ASPC..246...53P} accomplished this goal by obtaining simultaneous $V$ and $I$ band photometry. The survey has two units, one located in Las Campanas, Chile and the other in Haleakala, Maui. Each unit has two telescopes equipped with wide-field Minolta 200/2.8 APO-G telephoto lenses and a 2K$\times$2K Apogee CCD. Each telescope setup has an 8.8$^{\circ}\times8.8^{\circ}$ field of view. ASAS observed \thisstar{} in the $V$ band at 662 epochs at a median cadence of 2.04 days from UT 2001 January 31 to UT 2009 October 25. 

The eclipse signals were also recovered from observations by the All Sky Automated Survey for SuperNovae (ASAS-SN). Using two separate telescope units at Mount Haleakala in Hawaii and Cerro Tololo Observatory in Chile, ASAS-SN is monitoring the entire sky down to $V_\mathrm{mag} \approx$17 to detect new supernovae and transients \citep{2014ApJ...788...48S, 2017PASP..129j4502K}. Each unit has four 14\,cm aperture Nikon telephoto lens with 2K $\times$ 2K thinned CCDs, and can survey over 20,000 deg$^2$ each night, allowing the entire visible sky to be observed every 2 days. The telescope setup results in a 4.5$^{\circ}\times4.5^{\circ}$ field of view. The reduction pipeline is described in \citet{2017PASP..129j4502K}. ASAS-SN observed \thisstar{} 418 times at a mean cadence of 2.08 days from 2015 February 16 -- 2017 July 2 UT. 

We used the Box Least Squares algorithm \citep[BLS,][]{2002A&A...391..369K} to search for the period of the eclipse signal in the combined ASAS, ASAS-SN, and K2 data sets with the eclipse epoch fixed to be the center of the K2 event. The BLS spectrum has a strong detection at $72.416 \pm 0.016$ days. Uncertainties in this period estimation stem mainly from the difficulty in accurately determining the eclipse centroid.  We use this period estimation to guide further follow up observations. The period is also independently confirmed via a summed harmonics Fast Fourier Transform on the ensemble photometry, yielding a period of $72.428\pm0.030$ days, as well as via a Stellingwerf transform \citep{1978ApJ...224..953S} at a period of $72.417\pm0.030$ days.

To check for possible period changes, we fitted for a set of eclipse times over 1 year segments of the ASAS and ASAS-SN light curves, using the {\it K2} eclipse as a template. We find no eclipse timing variations, with a $2\sigma$ upper limit of $|\dot P| < 0.016$ day\,year$^{-1}$ over the 16 year baseline. From the description of the binary system evolution set out in Section~\ref{sec:evolution}, we expect no detectable period changes in the current system configuration, even if mass transfer is occurring at a rate of $10^{-8}\,M_\odot \, \mathrm{yr}^{-1}$ as inferred from the spectroscopy (Section~\ref{sec:sig_accretion}). Based on the simplest dimensional arguments using conservation of orbital angular momentum, we expect period changes of $\dot P/P \simeq 3 |\dot M|/M_{\rm don}$.  From that, we find that $\dot P$ is likely to be $\lesssim 4 \times 10^{-6}\,\mathrm{days}\,\mathrm{yr}^{-1}$, orders of magnitude smaller than our measurements could possibly detect.

\subsection{Ground-based photometric follow-up}
\label{sec:PHFU}

Following the period determination for the occultation of \thisstar{}, we targeted the predicted occultation in early 2017 September with a series of ground-based photometric observations. This occultation occurred five orbits after that observed by the {\it K2} mission, and was the first observable occultation post {\it K2} data release. 

Observations were obtained at the Hereford Arizona Observatory (HAO), using a 0.36\,m Meade LX200 GPS telescope with a Santa
Barbara Instrument Group (SBIG) ST-10XME CCD camera. We made observations during 16 nights between 2017 August 30 to 2017 September 19, successfully recording the ingress, full eclipse, egress, and three days of post-occultation baseline photometry. The observations and reductions procedure follows that described in \citet{2016MNRAS.458.3904R,2017arXiv170908195R}.  To check for color dependencies in the occultation depth, the observations were made in the $g'$ and $z'$ bands. On each night, continuous observations were performed over a two-hour span, with integration times of 20--40\,s. The per night average magnitudes are listed in Table~\ref{tab:lc_table} and plotted in Figure~\ref{fig:lc}.

We also obtained multi-band follow-up observations of the 2017 September occultation with the Perth Exoplanet Survey Telescope (PEST) located in Perth, Australia. PEST operates a fully automated 0.3\,m Meade LX200 telescope, coupled with a SBIG ST-8XME
CCD camera. The observations covered parts of the full eclipse and egress, and were made over 2017 September 2 -- 2017 September 12 period. To check for colour dependencies in the occultation, observations were made in the $B$, $V$, and $Ic$ bands. As with the HAO observations, the nightly averages of the light curves are shown in Figure~\ref{fig:lc}.

\begin{deluxetable*}{rrrrr}

\tablewidth{0pc}
\tabletypesize{\scriptsize}
\tablecaption{
        Differential photometry for \thisstar{}\tablenotemark{a} 
    \label{tab:lc_table}
}
\tablehead{
    \multicolumn{1}{c}{BJD}          &
    \multicolumn{1}{c}{Flux}            &
    \multicolumn{1}{c}{$\sigma$ Flux}            &
    \multicolumn{1}{c}{Instrument}         &
    \multicolumn{1}{c}{Filter}            
}
\startdata
2451940.8811800&0.9959&0.0290&ASAS&V\\
2451949.8930800&0.9922&0.0420&ASAS&V\\
2451953.8868900&0.9786&0.0310&ASAS&V\\
2451954.8783100&1.0144&0.0350&ASAS&V\\
2451961.8914600&0.9786&0.0330&ASAS&V\\
2451963.8718300&0.9886&0.0290&ASAS&V\\
2451965.8651100&1.0257&0.0300&ASAS&V\\
2451967.8676900&1.0125&0.0340&ASAS&V\\
2451978.8217600&0.9895&0.0400&ASAS&V\\
...&...&...&...&...\\
\enddata 
\tablenotetext{a}{
This table is available in a machine-readable form in the online journal. A portion is shown here for guidance regarding its form and content.
}

\end{deluxetable*}

\section{Spectroscopic observations}
\label{sec:spec}

We obtained spectroscopic observations of \thisstar{} with a series of facilities over 2017 August and September to measure the masses of the system, estimate spectral classifications, and to monitor for temporal changes in the spectroscopic features. These observations are summarized in Table~\ref{tab:specobssummary}.

A total of 11 observations were obtained using the Tillinghast Reflector Echelle Spectrograph (TRES) on the 1.5\,m telescope at Fred Lawrence Whipple Observatory, Mt. Hopkins, Arizona, USA. TRES is a fibre fed spectrograph with a spectral resolution of $R \equiv \lambda / \Delta \lambda = 44000$ over the wavelength range 3900--9100\,\AA{} via 51 echelle orders. Each observation is made of three sequential exposures, combined to minimize the impact of cosmic rays, and were reduced as per \citet{2010ApJ...720.1118B}. Our observations spanned the period of 2017 July 31 -- 2017 September 25, over orbital phases between 0.25 and 0.75. Of the 11 observations, 5 were obtained during the occultation at phase 0.5.

We observed \thisstar{} six times between 2017 August 28 and 2017 September 15 with the 2.7 m Harlan J.\ Smith Telescope and its Robert G.\ Tull Coud\'e Spectrograph \citep{1995PASP..107..251T} at McDonald Observatory, Mt. Locke, Texas, USA. We used the TS23 spectrograph configuration, providing $R=60,000$ over 58 echelle orders. The wavelength coverage is 3570-10200 \AA, and is complete below 5691 \AA\ with increasingly large inter-order gaps red-ward of this point; notably, H$\alpha$ falls into one of these gaps and is not captured. We obtained one spectrum per epoch. The data were reduced and the spectrum extracted and wavelength calibrated using standard IRAF tasks.

We also obtained an observation of \thisstar{} using the 2.4\,m Automated Planet Finder (APF) and the Levy spectrograph, located at Lick observatory, Mt Hamilton, California, USA. The APF is coupled with a high resolution, slit-fed, spectrograph that works at a typical resolution of R $\approx$ 110,000. The telescope is operated by a dynamic scheduler, has a peak overall system throughput of 15\%, and its data reduction pipeline extracts spectra from 3750-7600\AA \, \citep{Vogt2014, Burt2015}. The APF spectrum used in this work was extracted from a single, 20 minute exposure taken with the 1x3" slit on 2017 July 26.

\begin{deluxetable*}{llrrrr}
\tablewidth{0pc}
\tabletypesize{\scriptsize}
\tablecaption{
    Summary of spectroscopic observations\label{tab:specobssummary}
}
\tablehead{
    \multicolumn{1}{c}{Telescope/Instrument} &
    \multicolumn{1}{c}{Date Range}          &
    \multicolumn{1}{c}{Number of Observations} &
    \multicolumn{1}{c}{Resolution}          &
}
\startdata
APF 2.4\,m & 2017 Jul 26 & 1 & 110000\\
FLWO 1.5\,m/TRES & 2017 Jul 31 -- 2017 Sep 25 & 11 & 44000 \\
McDonald 2.7\,m & 2017 Aug 28 -- 2017 Sep 15 & 6 & 60000 \\
\enddata 
\end{deluxetable*}

\subsection{Radial velocities of the binary}
\label{sec:RV}

\begin{figure} [ht]
\plotone{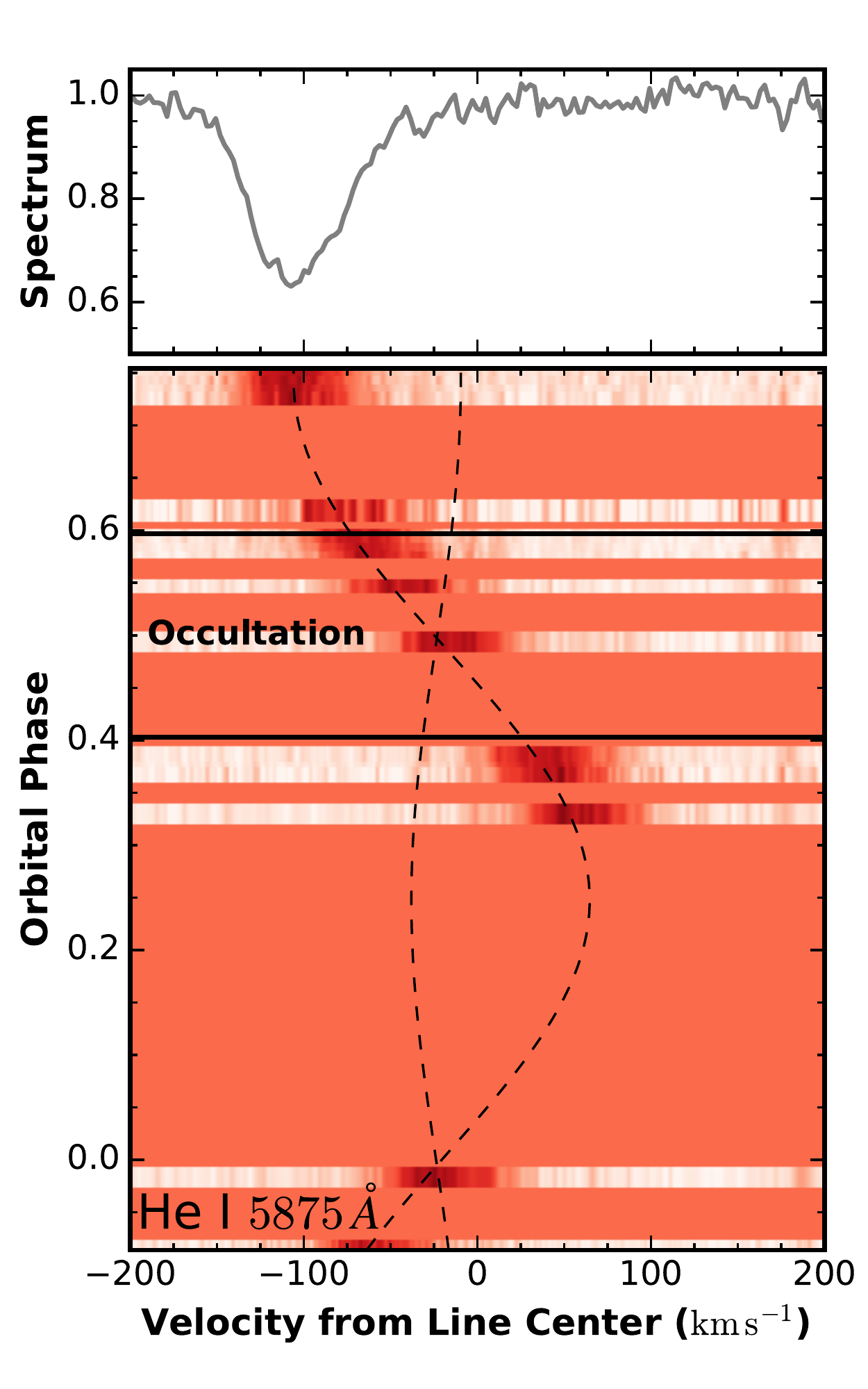}
\caption{
He I lines trace the velocity of the donor star. The \textbf{top panel} shows the He I line at the end of our observation campaign, at an orbital phase of 0.75. The \textbf{bottom panel} shows the Doppler map of the He I lines over the orbital phase of the \thisstar{} system, ordered such that the latest observation (top row) corresponds to the spectrum shown on the top panel. The color gradient represents the depth of the He absorption line. Velocity with respect to the rest frame He line is plotted on the horizontal axis, temporal variations of the spectrum plotted along the vertical axis. The Keplerian orbital fit for both stellar components are marked by the dashed lines. Note the He I lines are present only in the spectrum of the donor, indicative of it being a hot B-star. 
\label{fig:He}}
\end{figure}

The spectra of \thisstar{} show obvious signs of blending by the two stellar components in the system. Luckily, only the donor star is hot enough to exhibit strong He I absorption lines. As such, we make use of the unblended He I lines at 5876\,\AA{} and 6678\,\AA{} to measure the radial velocity of the donor star (Figure~\ref{fig:He}). To measure the velocity of the accretor, we perform a least-squares deconvolution of the spectrum to derive the broadening kernel of the spectral lines. The broadening kernel contains the radial velocity information of the system, since the observed spectrum is the convolution of the kernel of the two stars and a synthetic non-rotating single star spectral template. Following the technique laid out in \citet{1997MNRAS.291..658D}, a deconvolution is performed between the observed spectrum and a single-star non-rotating synthetic spectral template. The template is generated with the SPECTRUM code\footnote{\texttt{http://www1.appstate.edu/dept/physics/spectrum/spectrum.html}} \citep{1994AJ....107..742G} with the ATLAS9 model atmospheres \citep{2004astro.ph..5087C}, over the wavelength range of 4000-6100\,\AA, and has the atmospheric properties of a $T_\mathrm{eff}=12000$\,K, $\logg = 4.5$ main-sequence star. We find that broadening profiles from least-squares deconvolutions often resolve blended lines better than classical cross correlation functions, due to the sharp-edge nature of rotational broadening kernels. 

\begin{figure} [ht]
\plotone{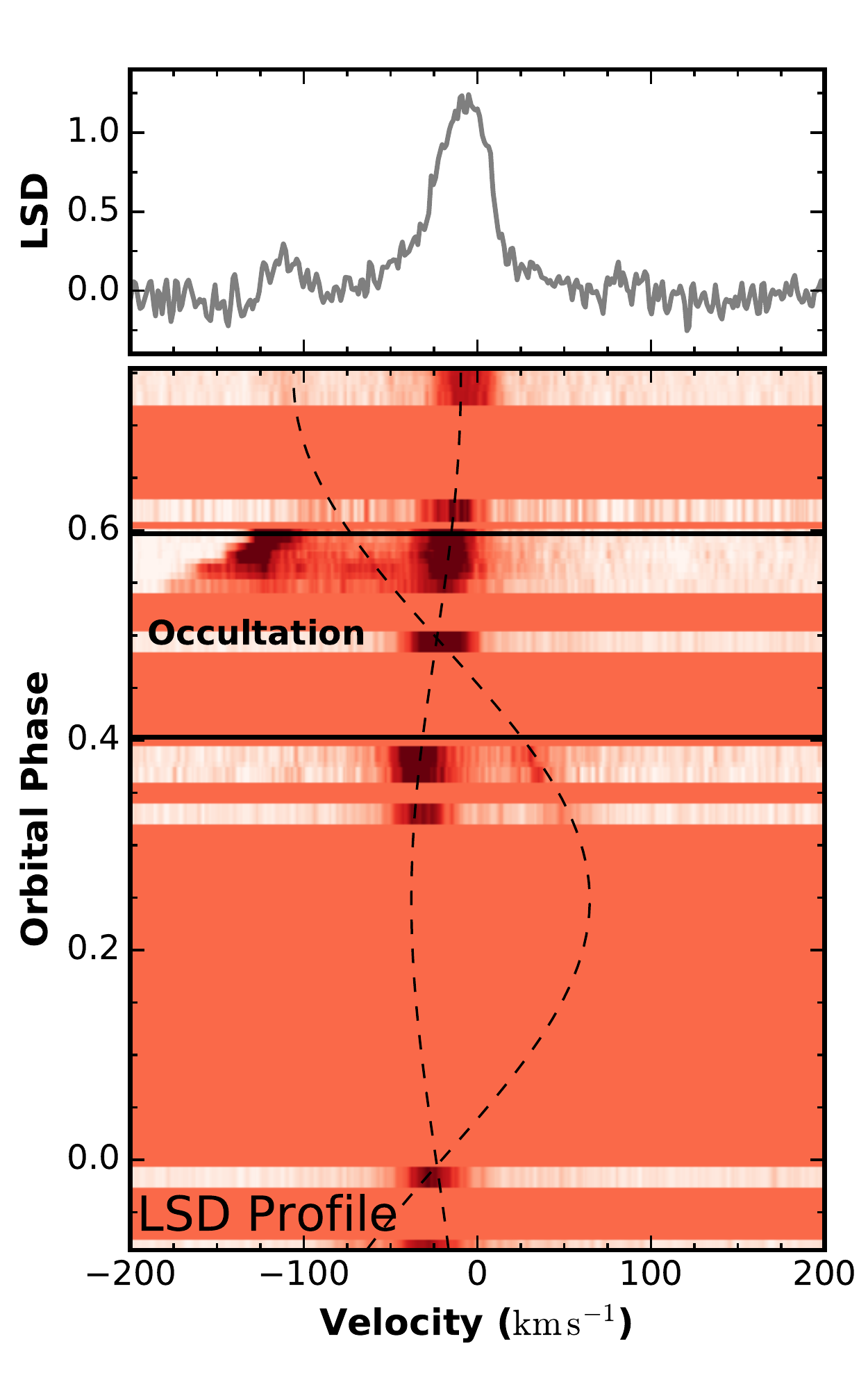}
\caption{
Doppler tomographic maps of the line broadening kernel variations are plotted. The line broadening profiles are derived from a least-squares deconvolution (LSD) of the observed spectra against a non-rotating synthetic spectral template over the wavelength region of 4000--6100\,\AA, and represent line broadening kernels for the photospheric lines in both stars. Similar to Figure~\ref{fig:He}, the \textbf{top panel} shows the line profile at an illustrative epoch, while the \textbf{bottom panel} shows the temporal variations in a color map, with each row representing an observation epoch, and top row corresponding with the example profile plotted on the top panel. The Keplerian orbits of both stellar components are marked by the dashed lines, and the spectral contributions of both stellar components can be identified in the line broadening kernels. Note the absorption features from the accretion disk that are seen during eclipse, blue shifted with respect to the accreting star during egress. These are transmission spectroscopic features that mark the Keplerian velocity of the sections of the gas disk occulting the donor star. 
\label{fig:lsd}}
\end{figure}

\begin{figure} [ht]
\plotone{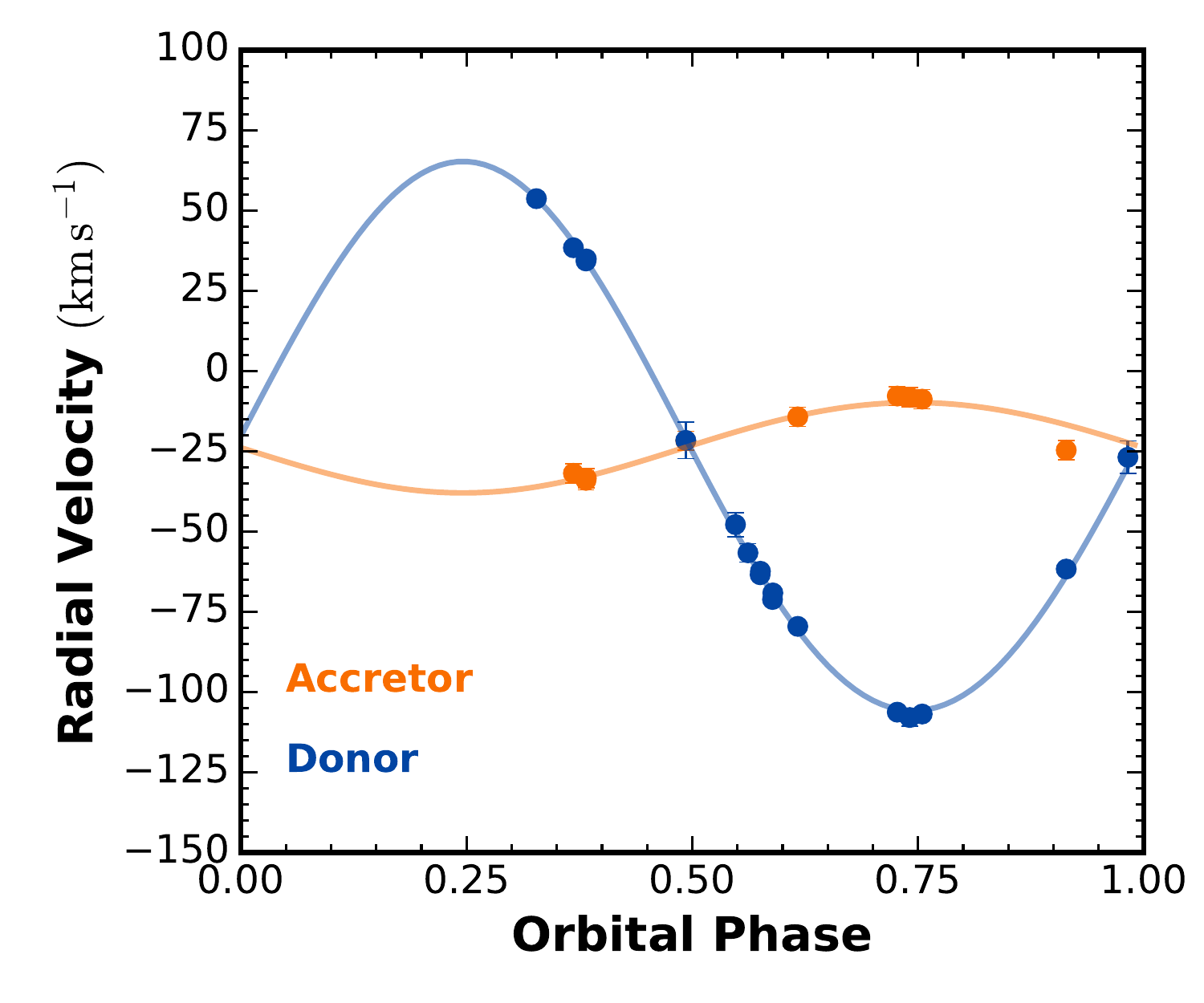}
\caption{
Radial velocities of the accretor (orange) and donor (blue) are shown. The occultation is centred at phase 0.5. The best fit Keplerian orbits are marked by the solid lines. 
\label{fig:rv}}
\end{figure}

Temporal variations of the broadening profile, as a function of the orbital phase, are shown in Figure~\ref{fig:lsd}. These temporal variations capture the radial velocity orbit of the system, and have the potential of revealing line width and asymmetry variations if they are present. We fit the broadening profiles with a two-star model, and each kernel of each star is modeled with the convolution of a rotational kernel and a macroturbulence kernel, along with the parameters for velocity centroid and height ratio. The velocity of the broadening kernel of the donor is fixed to that measured from the He I line centroids, while their broadening values are fitted for simultaneously with the fitting of the donor's profile. The fitting is performed via a Markov chain Monte Carlo (MCMC) exercise, using the {\it emcee} affine invariant ensemble sampler \citep{2013PASP..125..306F}. The derived radial velocities are presented in Table~\ref{tab:rv_table}, and plotted in Figure~\ref{fig:rv}.

Keplerian fits to the radial velocities yield the dynamical masses of the system. The Keplerian orbit parameters and associated uncertainties are determined via an MCMC analysis. To yield realistic uncertainty estimates and to account for underestimation of per-point velocity measurement errors, the per-point velocity uncertainties were inflated such that the reduced $\chi^2$ for the best fit solution after the MCMC burn-in chain is at unity. We did not attempt to correct the velocities from different facilities for systematic offsets. Instead we chose to increase the formal uncertainties to account for this effect. The per-point uncertainties were assigned to be at least 0.5\,\kms{} to account for these possible systematic uncertainties. We find, for assumed circular orbits, that the radial velocity amplitudes are $K_1 = \Kone$\,\kms{} and $K_2 = \Ktwo$\,\kms{}, resulting in minimum mass measurements of $M_\mathrm{acc} = \Mone\,M_\odot$ and $M_\mathrm{don} = \Mtwo\,M_\odot$. The residuals from the best fit model have scatters of 3.0\,\kms{} and 1.4\,\kms{} for the accretor and donor velocities, respectively. To test the effect of combining velocities from multiple instruments in our analysis, we recomputed the masses from the TRES velocities alone, finding $M_\mathrm{acc} = 3.14\pm0.14 \, M_\odot$ and  $M_\mathrm{don} = 0.54\pm0.02 \, M_\odot$, consistent with the combined analysis. We also performed an independent radial velocity analysis of the TRES spectra via the two-dimensional cross correlation technique {\it TODCOR} \citep{1994ApJ...420..806Z}, finding $M_\mathrm{acc} = 3.01 \pm 0.17 \, M_\odot$, and $M_\mathrm{don} = 0.508 \pm 0.017 \, M_\odot$, consistent with the masses quoted above to within $1\sigma$. 

We checked that the orbit is indeed consistent with being nearly circular, with eccentricity formally constrained to be $e = 0.021 \pm 0.010$. While we cannot rule out a small, but non-zero eccentricity for the system, significantly more radial velocities are required to avoid biases at near-zero eccentricities \citep[e.g.][]{1971AJ.....76..544L,1978Obs....98..122B}. We note, however, that non-zero eccentricities have been detected for other Algols \citep[e.g. TT Hydrae][]{2007ApJ...656.1075M}, so such follow-up observations are worthwhile.

\begin{deluxetable*}{lrrrrr}
\tablewidth{0pc}
\tabletypesize{\scriptsize}
\tablecaption{
       Relative radial velocities
    \label{tab:rv_table}
}
\tablehead{
    \multicolumn{1}{c}{BJD}          &
    \multicolumn{1}{c}{RV$_1$}\tablenotemark{a}             &
    \multicolumn{1}{c}{$\sigma$ RV$_1$}      &
    \multicolumn{1}{c}{RV$_2$}\tablenotemark{b}         &
    \multicolumn{1}{c}{$\sigma$ RV$_2$} \tablenotemark{c}           & 
    \multicolumn{1}{c}{Inst}            \\
    \multicolumn{1}{c}{(TDB)} &
    \multicolumn{1}{c}{$(\mathrm{km\,s}^{-1})$} &
    \multicolumn{1}{c}{$(\mathrm{km\,s}^{-1})$} &
    \multicolumn{1}{c}{$(\mathrm{km\,s}^{-1})$} &
    \multicolumn{1}{c}{$(\mathrm{km\,s}^{-1})$} &
}
\startdata
2457960.7368 & -24.62 & 0.21 & -61.67 & 0.50 & APF \\
2457965.6830$^d$ & ... & ... & -26.84 & 1.79 & TRES \\
2457990.6635$^d$ & ... & ... & 53.73 & 0.55  & TRES \\
2457993.6303 & -31.83 & 0.15 & 38.44 & 0.50 & McDonald \\
2457994.6378 & -34.05 & 0.31 & 34.32 & 0.50 & McDonald \\
2457994.6673 & -33.38 & 0.23 & 34.94 & 0.50 & TRES \\
2458002.6414 & -21.70 & 0.33 & -21.58 & 2.00 & TRES \\
2458006.6275$^e$ & ... & ... & -47.81 & 1.31 & TRES \\
2458007.6279$^e$ & ... & ... & -56.62 & 1.02 & McDonald \\
2458008.6041$^e$ & ... & ... & -63.44 & 0.50 & McDonald\\
2458008.6290$^e$ & ... & ... & -62.36 & 0.50 & TRES \\
2458009.5981$^e$ & ... & ... & -71.15 & 0.50 & McDonald \\
2458009.6244$^e$ & ... & ... & -69.11 & 0.50 & TRES \\
2458011.6228 & -14.25 & 0.16 & -79.53 & 0.50 & McDonald \\
2458019.6026 & -7.77 & 0.16 & -106.26 & 0.50 & TRES \\
2458020.6046 & -8.17 & 0.30 & -107.93 & 0.96 & TRES \\
2458021.6009 & -8.74 & 0.10 & -106.84 & 0.50 & TRES \\
\enddata 
\tablenotetext{a}{
Velocity of the accretor measured from the line broadening kernels.
}
\tablenotetext{b}{
Velocity of the donor measured from the He I lines, which are present only in the donor spectrum. 
}
\tablenotetext{c}{
A minimum uncertainty of 0.50\,\kms{} is adopted to account for systematic uncertainties. 
}
\tablenotetext{d}{
Only velocities of the donor were measured; could not disentangle the velocities of the accretor and donor from the line broadening kernels.
}
\tablenotetext{e}{
Only velocities of the donor were measured; the photospheric lines of the accretor were contaminated by the disk absorption lines.
}
\end{deluxetable*}

\subsection{Spectral classification}
\label{sec:spec_class}

Stellar classification was particularly difficult given the complexity of the system: the spectra are blended, both stars exhibit signatures of chemical peculiarity (see Section~\ref{sec:chemical_peculiarity}), active accretion prevents the use of Balmer line strengths for temperature estimates, and both stars are severely reddened by interstellar and circumstellar material. As such, no single set of synthetic template spectra and fluxes fit the observed spectra well. 

\begin{figure*} [ht]
\begin{tabular}{cc}
\includegraphics[width=0.5\linewidth]{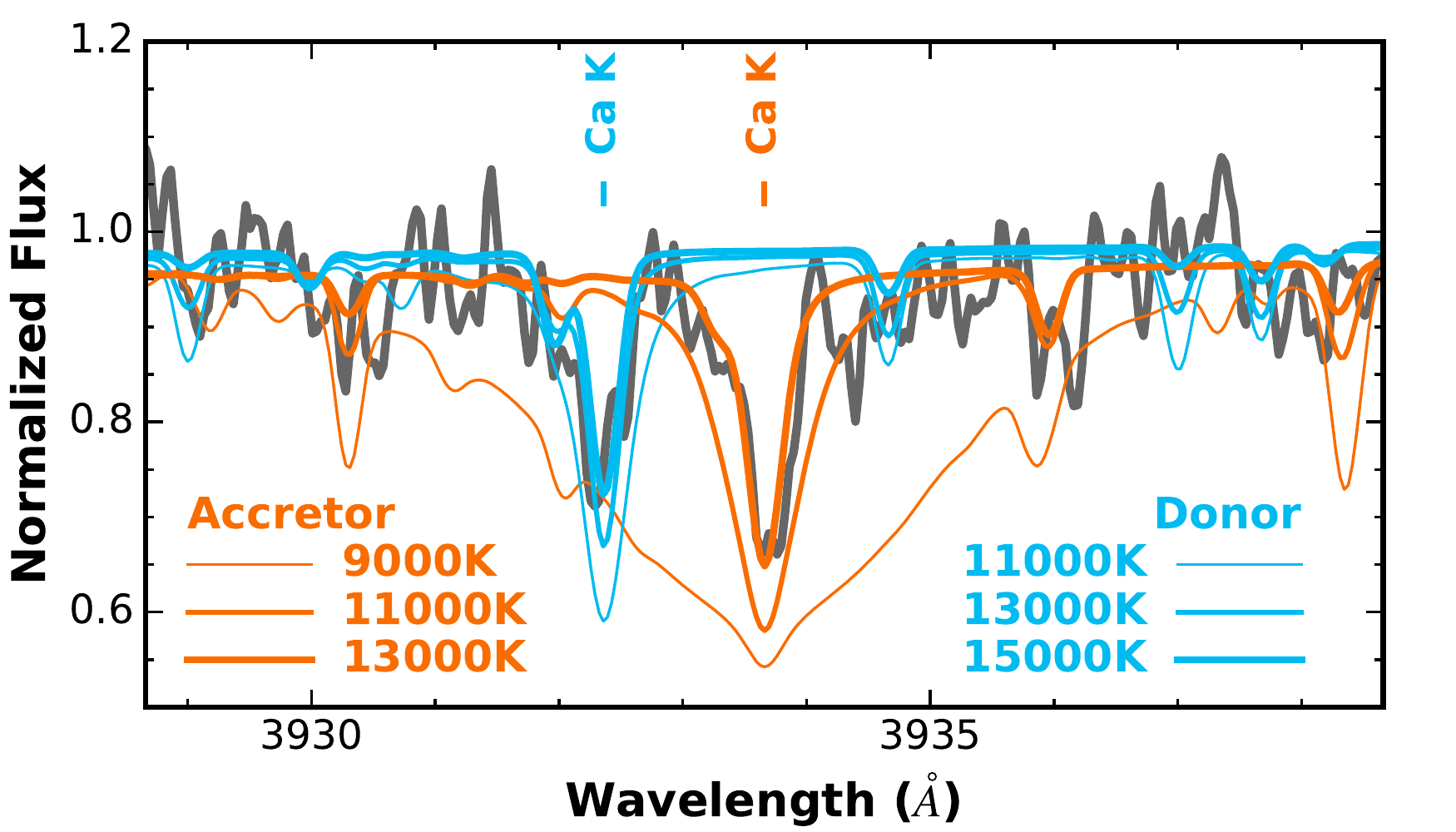} &
\includegraphics[width=0.5\linewidth]{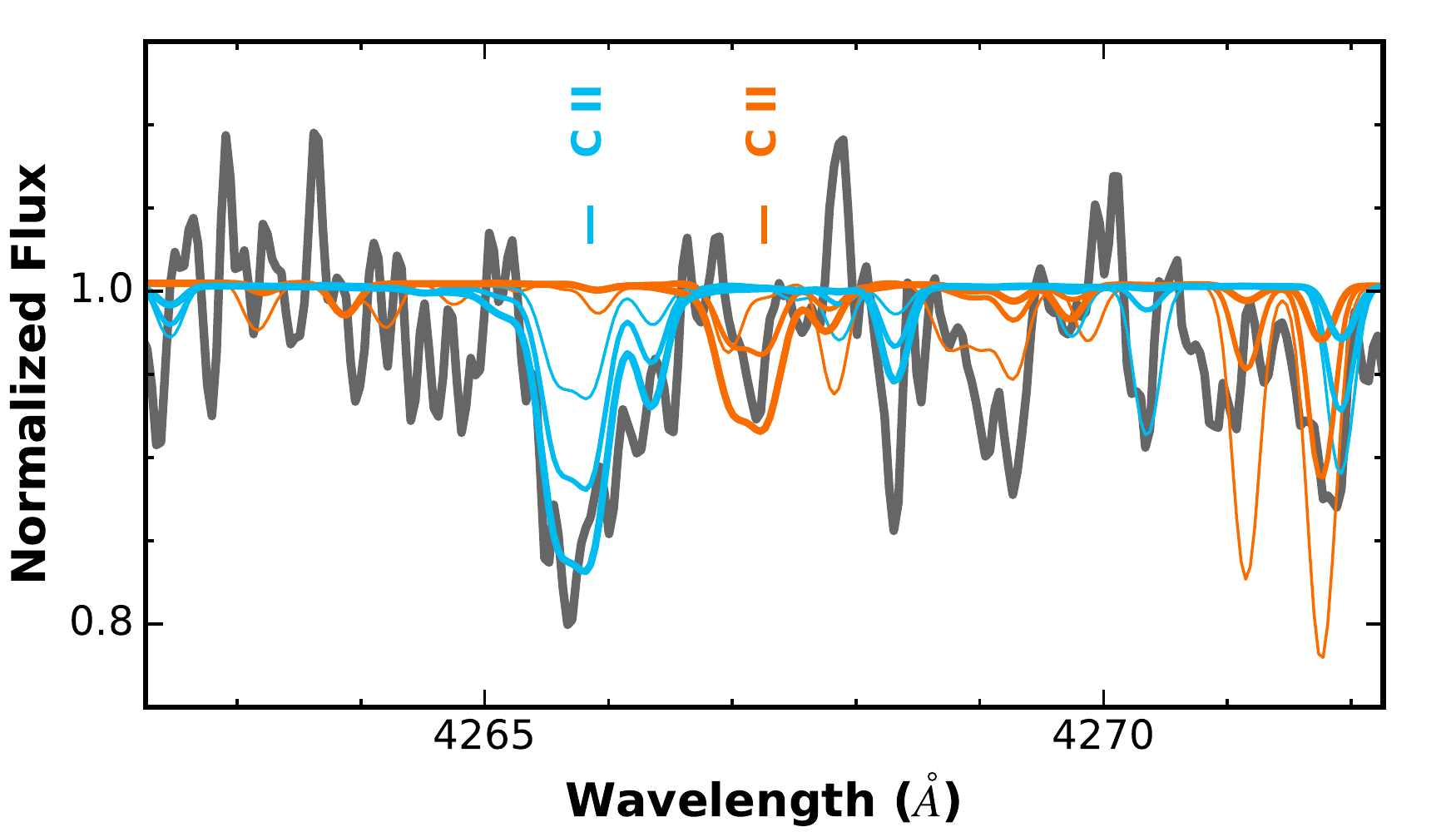} \\
\end{tabular}
\caption{
Ca II K and C II lines are temperature sensitive along the A--B spectral range. We compare the strengths of these lines against synthetic templates from ATLAS12 \citep{2005MSAIS...8..189K}. The observed spectrum is shown in grey, and is an average spectrum over 3 nights at orbital phase of 0.25. The spectrum has been shifted to the rest velocity of the accreting star. The 9000\,K, 11000\,K, and 13000\,K templates are shown in orange for the accreting star, templates of 11000\,K, 13000\,K, and 15000\,K, shifted to the velocity of the donor star, are shown in blue. 
\label{fig:spec_class}}
\end{figure*}

Since the stars exhibit chemical peculiarity and accretion signatures, we did not perform a global spectral matching of observed and synthetic spectra. Instead, we make use of temperature sensitive spectral features. For example, strong and broad near-UV/optical He I absorption lines are prominent features in B-star spectra, but decrease in strength toward cooler temperatures.  As a result, He I lines can be useful temperature indicators. Figure~\ref{fig:He} shows the radial velocity orbital phase variations of the donor star by its He I absorption feature. Note that we see He I only at the spectrum of the donor star, not the accretor, suggesting that the donor star is hotter than $\approx 13000\,\mathrm{K}$. The Ca II H \& K lines grow weaker and narrower for earlier type A and B stars, while C II lines grow in strength for B stars. Figure~\ref{fig:spec_class} shows the observed spectra of \thisstar{} for the Ca II K and C II lines, compared to synthetic spectra from ATLAS12 models \citep{2005MSAIS...8..189K} taken from the POLLUX spectral database \citep{2010A&A...516A..13P}. Synthetic spectra with metallicity of $\mathrm{[Fe/H]} = +0.5$ were chosen, since the stellar surface of both stars appears metal enriched and chemically peculiar (Section~\ref{sec:chemical_peculiarity}). We also adopt models with a surface gravity of $\logg = 4.5$ for the accretor, corresponding to its expected properties after mass transfer (see Section~\ref{sec:evolution}), and a gravity of $\logg = 3.5$ for the donor, the lowest gravity template available in the ATLAS12 grid. We take as the `observed spectrum' to be evaluated an average of the spectra taken over 2017 September 23 -- 25, at orbital phase 0.75--0.77, where the radial velocity separation between the accretor and donor were at its greatest. We find the accretor star to be consistent with an A0 spectral type with an effective temperature of $12000\pm1000$\,K, while the donor is likely a B7 star, with an effective temperature of $15000\pm1000$\,K. The similar temperatures between the accretor and donor star are consistent with the grey eclipse observed during our follow-up observations (Section~\ref{sec:PHFU}).

Broadening of the spectral lines informs us of the projected rotational velocity of the stars. We fit for rotation profiles \citep[defined in][]{2005oasp.book.....G} to the least-squares deconvolution kernels taken at phase 0.25 and 0.75, where the velocities of the two stars are well separated. We find both stars exhibit similar line broadenings, with the accretor rotating at $v\sin i = 25.6\pm4.5\,\kms$, and the donor star with $v\sin i = 31\pm11\,\kms$. We note, however, that our quoted rotational broadening velocities do not account for the effects of macroturbulence. The broadening profile of both stars in the \thisstar{} system deviate significantly from the standard rotational profiles, and the fits that incorporate both \vsini{} and macroturbulence are highly degenerate. A similar issue was encountered when analysing the spectra of $\epsilon$ Aurigae, with multiple \vsini{} values quoted, ranging from $5\,\kms$ \citep{2011A&A...530A.146C} to 41\,\kms{} \citep{2010AJ....139.1254S}. Regardless of the assumptions on non-rotational line broadenings, the stars in both \thisstar{} and $\epsilon$ Aurigae are rotating significantly slower than the average A and B stars \citep[e.g.][]{2007A&A...463..671R}, and certainly not near breakup velocity as expected for actively accreting systems. \citet{2010MNRAS.406.1071D} suggest a strong stellar wind as a mechanism for angular momentum dumping. They expect that strong differential rotation, induced by accretion, can generate strong magnetic fields that enhance the stellar wind and mass loss, spinning down the stars. 

\begin{figure*} [ht]
\includegraphics[width=0.95\linewidth]{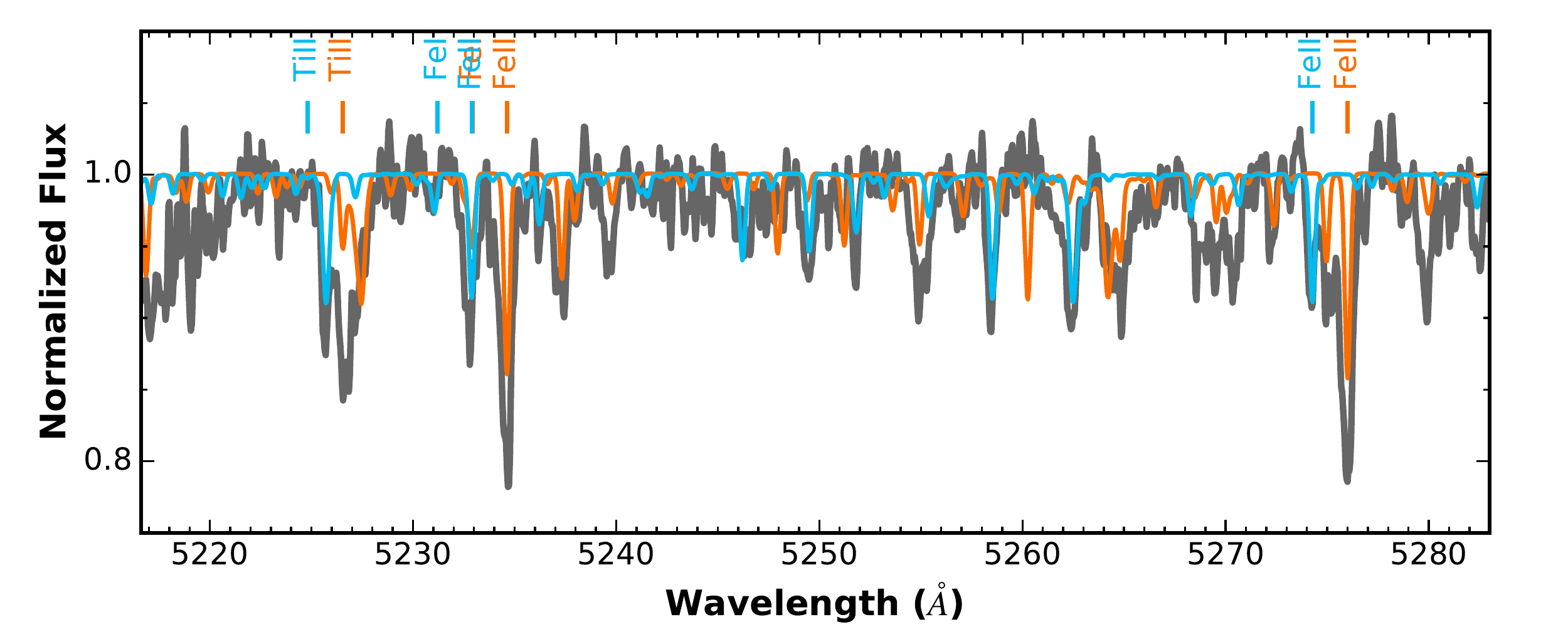} 
\caption{
Relative line strengths in the observed composite spectrum can be used to determine the flux ratio of the system. An illustrative portion of the spectrum is plotted, with the observations in grey, A 12000\,K ATLAS12 model spectrum in orange for the accretor, and a 15000\,K model spectrum in blue, shifted in velocity to match the donor star. Strong lines are labelled.
\label{fig:fluxratio}}
\end{figure*}

The relative line strengths of the accretor and donor spectra also inform us of the flux ratio of the system. To measure the flux ratio, we fit for the contributions of the donor and accretor stars over the entire TRES spectrum, order by order, using the ATLAS12 models. Only a few orders have high signal-to-noise and enough lines from both stars to constrain the flux ratio. We find the spectral region over 48 orders between 4000--6000\,\AA{} satisfy this requirement well, yielding $F_\mathrm{acc} / (F_\mathrm{don} + F_\mathrm{acc}) = 0.52 \pm 0.18$, with the uncertainty being the scatter in the flux ratios measured. An illustrative section of the spectral fitting is shown in Figure~\ref{fig:fluxratio}. 

\begin{figure} [ht]
\plotone{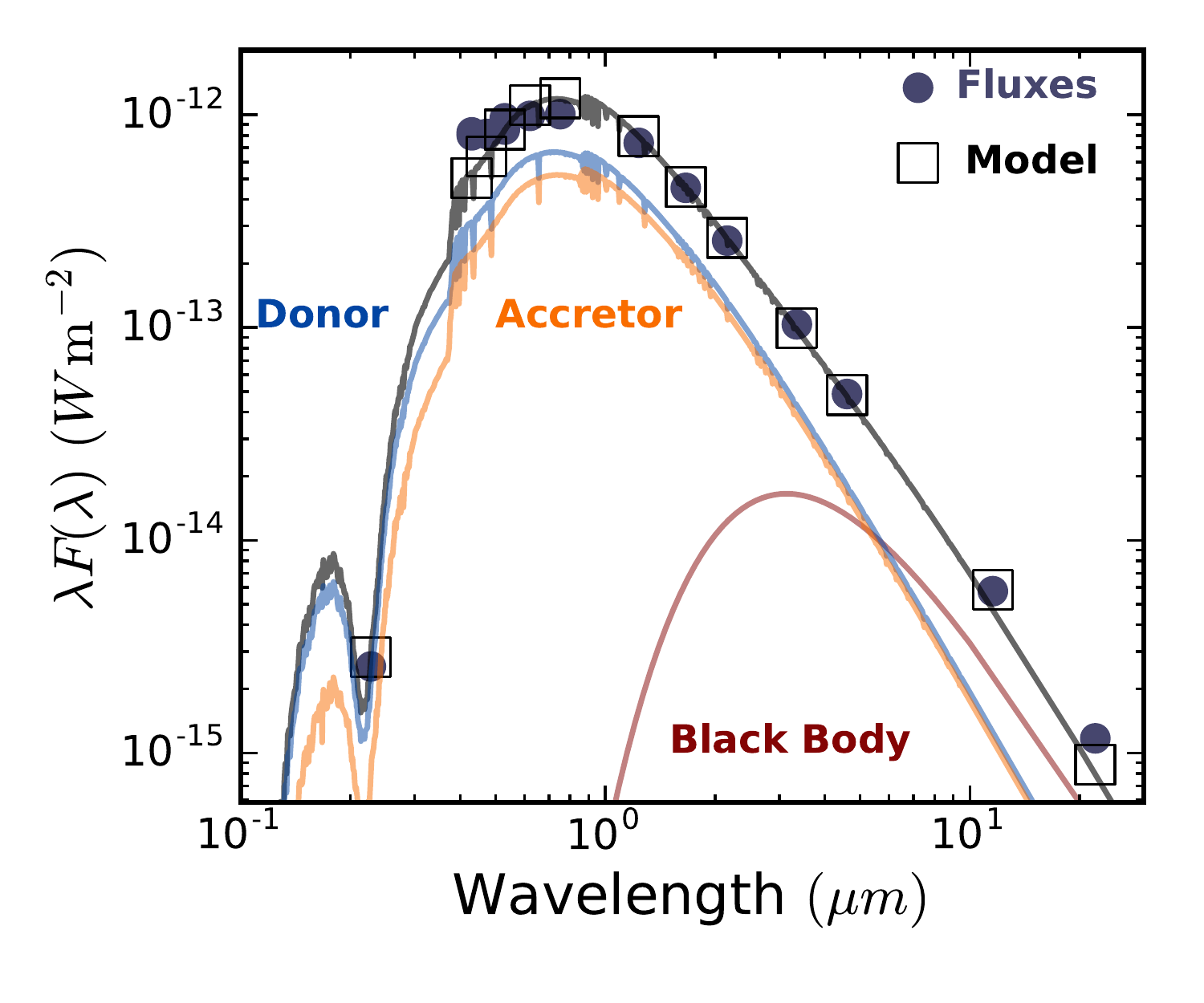}
\caption{
The SED of \thisstar{} is plotted. The measured fluxes are marked by the filled grey circles, while the best fit model fluxes are marked by the open squares. The SED is fitted with a 12000\,K (accretor, orange) and a 15000\,K (donor, blue) model atmospheres from \citet{2004astro.ph..5087C}. We find a best fit reddening of $E(B-V)=1.07$, consistent with significant interstellar reddening along the line of sight and circumstellar reddening around the system. We also find minimal IR excess in the WISE bands, which can be modeled well by a 1200\,K black body, shown in red. \label{fig:sed}}
\end{figure}

Reddening, UV, and IR excess are constrained by the spectral energy distribution (SED, Figure~\ref{fig:sed}). We model the SED with 12000\,K and 15000\,K synthetic templates from \citet{2004astro.ph..5087C}, and fit for a black body component to constrain any IR excess, and an extinction coefficient to account for interstellar and circumstellar reddening. Since both stellar components have very similar temperatures, the flux ratio is impossible to constrain using the SED only, so we fix the flux ratio to 0.52 as determined from spectral line fitting. We recover significant reddening of $E(B-V) = 1.07$, and is better fit with an extinction law of $R(V) = A(V) / E(B - V) = 2.5$. Minimal IR excess is detected in the SED, and can be rectified by a black body of 1200\,K. The IR excess may be due to contaminating M-dwarfs along the line of sight, or low levels of a dusty envelope around either of the stars. 

\subsection{Chemical peculiarity}
\label{sec:chemical_peculiarity}

\begin{figure*} [ht]
\centering
\includegraphics[width=0.8\linewidth]{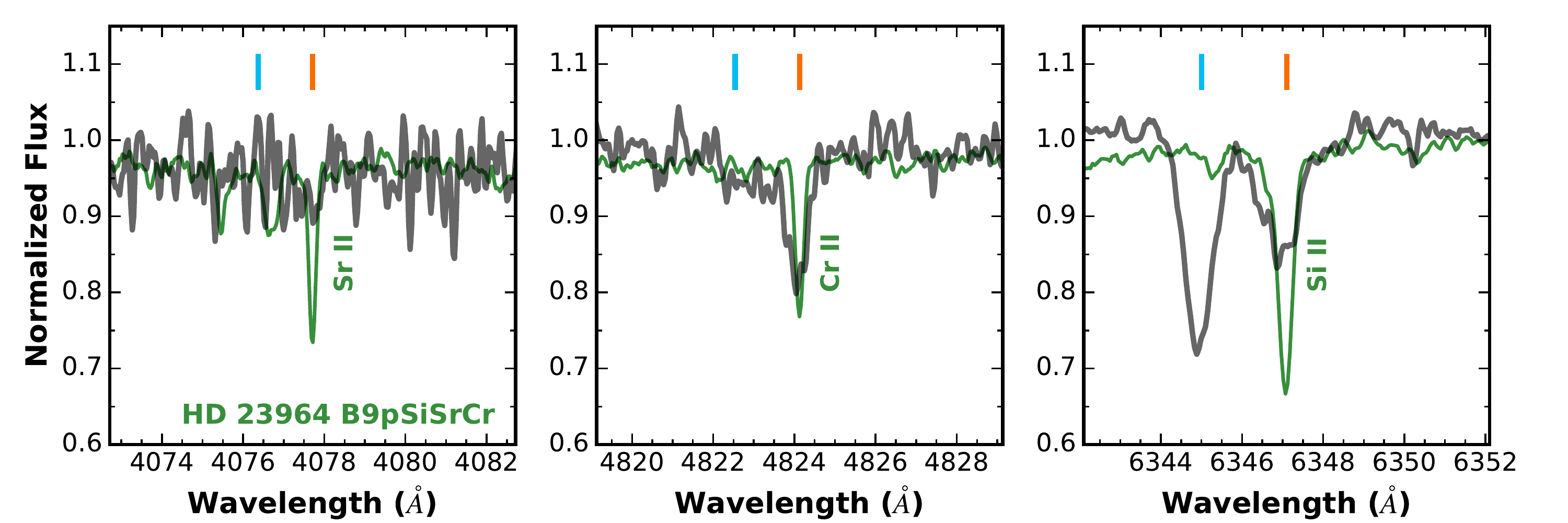}
\caption{
The presence of Si and Cr are indicators of chemical peculiarity in A and B stars. We compare the spectrum of \thisstar{} against the TRES spectrum of a known chemically peculiar Bp star enriched in Sr, Cr, and Si \citep[identified from the catalog compiled by][]{2009A&A...498..961R}. Chemical enhancement in Cr is seen in the accretor star but Si is enhanced in both the donor and accretor. In contrast, the 4077\,\AA{} Sr II lines are weak in both the donor and accretor. We therefore classify the donor star as a B9p Si star, and the accretor as an A0p Si Cr star. The spectrum of the known Bp star is plotted in green, with the specific lines labelled. The spectrum of \thisstar{} is plotted in grey. The expected wavelength of the lines of interest are labelled by the orange vertical mark for the accretor, and blue for the donor. 
\label{fig:SiSrCr}}
\end{figure*}

\begin{figure*} [ht]
\centering
\includegraphics[width=0.6\linewidth]{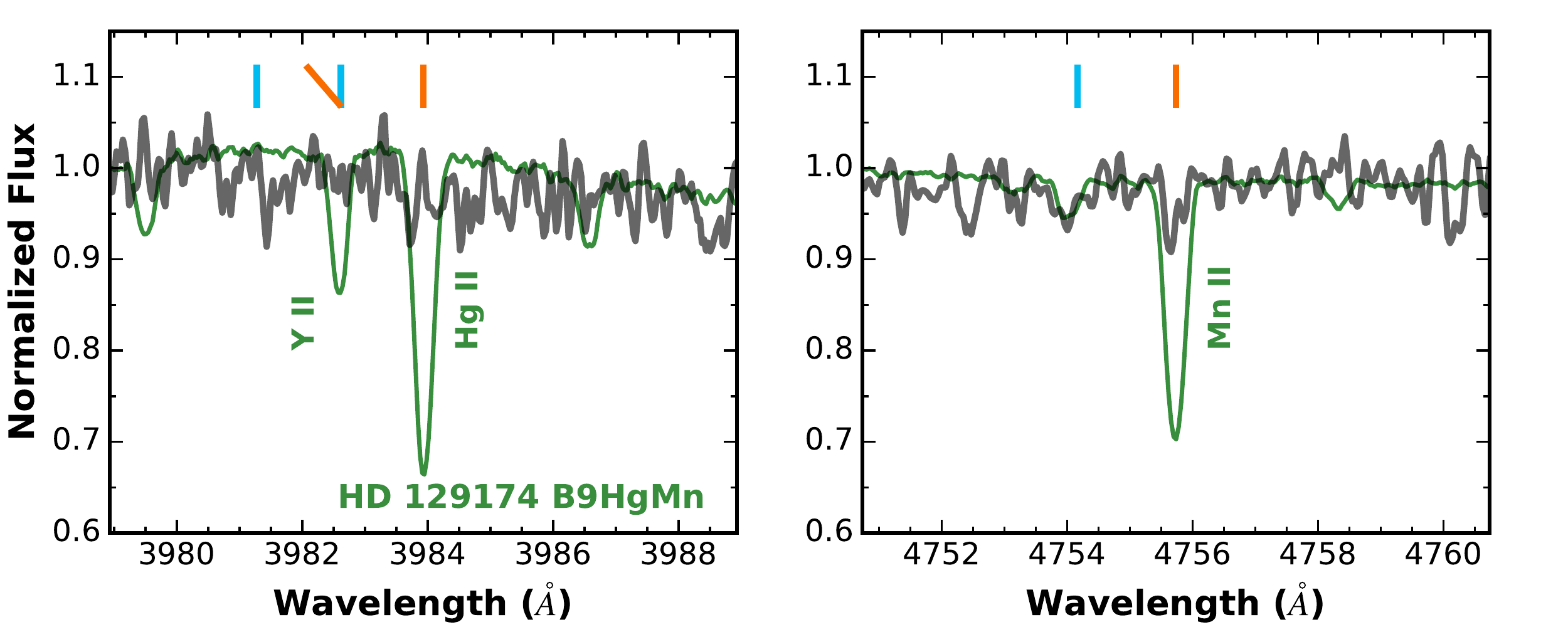}
\caption{
A class of chemically peculiar A and B stars are enriched in Hg, Mn, and Y. We compare the spectrum of \thisstar{} against the TRES spectrum of a B9p Hg Mn star from \citet{2009A&A...498..961R}. We see no enhancements in Hg and Mn in either the accretor or donor stars. The spectrum of the known Bp star is plotted in green, with the specific lines labelled. The spectrum of \thisstar{} is plotted in grey. The expected wavelength of the lines of interest are labelled by the orange vertical mark for the accretor, and blue for the donor. 
\label{fig:HgMn}}
\end{figure*}

Some A and B stars exhibit chemical and magnetic peculiarity in their atmospheres. The spectra of both stars in the \thisstar{} system are consistent with classification as chemically peculiar stars.  In particular, Figure~\ref{fig:SiSrCr} shows that both the donor and accretor star exhibit signatures of strong Si enhancement. Enhancement in the Cr lines are also seen in the accretor star, but not the donor star. We did not see enhancement in Sr, which is common amongst stars that are enriched in Si and Cr. We also searched for enhancements in Y, Hg, and Mn (Figure~\ref{fig:HgMn}), seen in the HgMn class of Bp/Ap stars, but did not detect their presence. The slow rotational ($\la$ 30 km s$^{\rm -1}$) velocities of both stars in the \thisstar{} system are consistent with the presence of strong magnetic fields, and both stars reside within the roughly B6-F4 spectral range of known Bp/Ap stars \citep[e.g.][]{1981ApJ...244..221W,1996Ap&SS.237...77S}.  Therefore, we have classified the donor and accretor stars in \thisstar{} as BpSi and ApSiCr stars, respectively.

\subsection{Spectroscopic signatures of accretion}
\label{sec:sig_accretion}

\begin{figure*} [ht]
\begin{tabular}{ccc}
\includegraphics[width=0.3\linewidth]{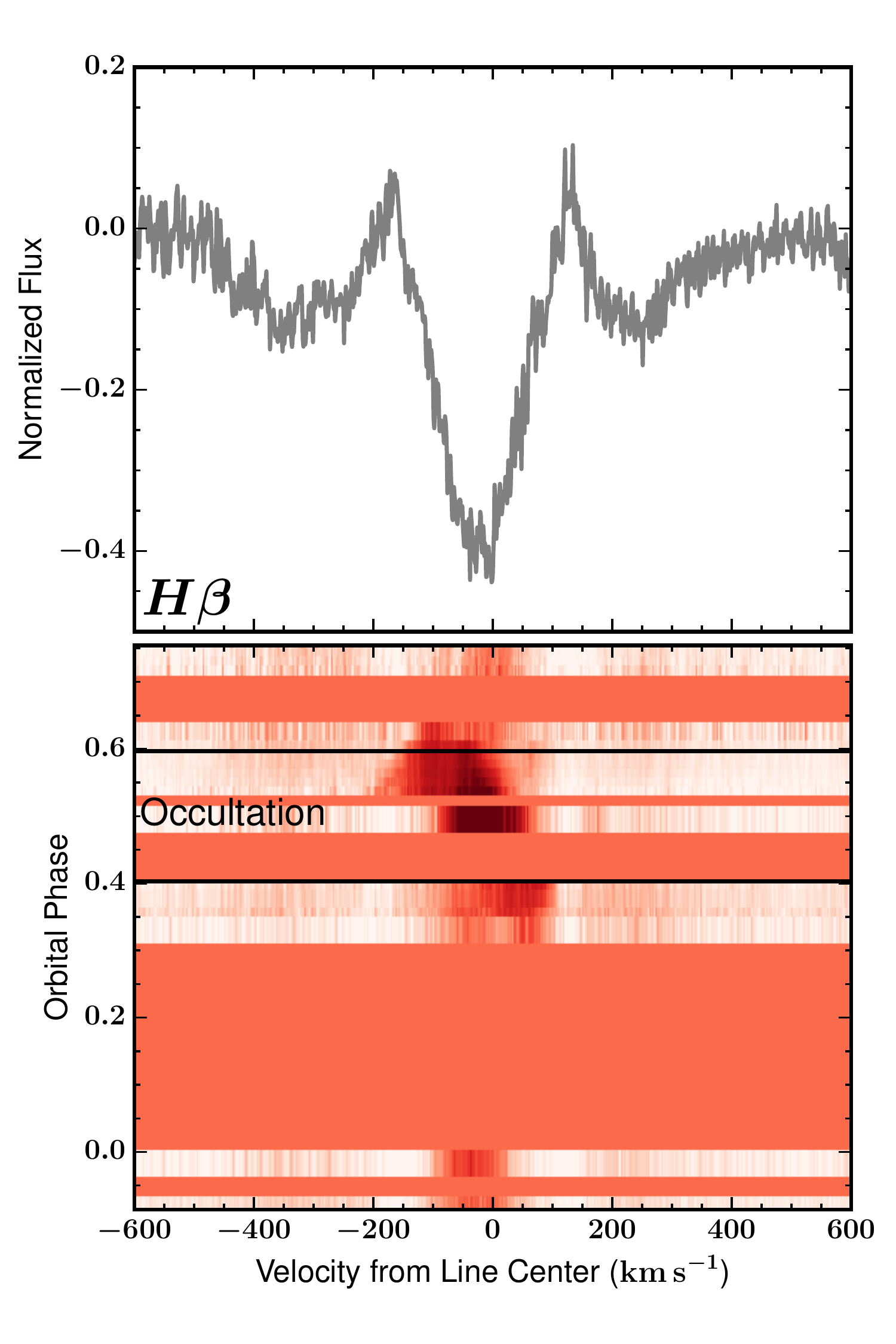} &
\includegraphics[width=0.3\linewidth]{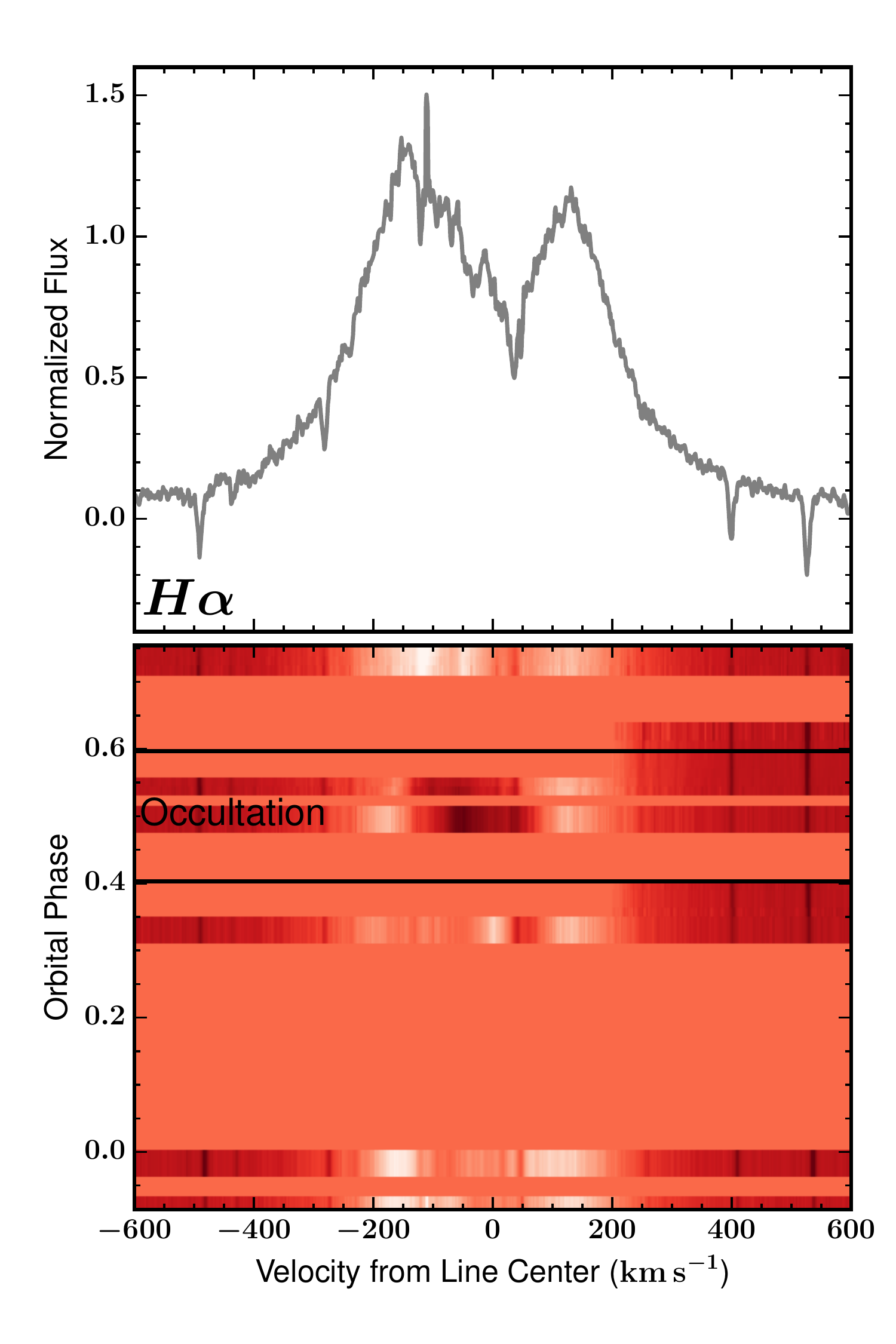} &
\includegraphics[width=0.3\linewidth]{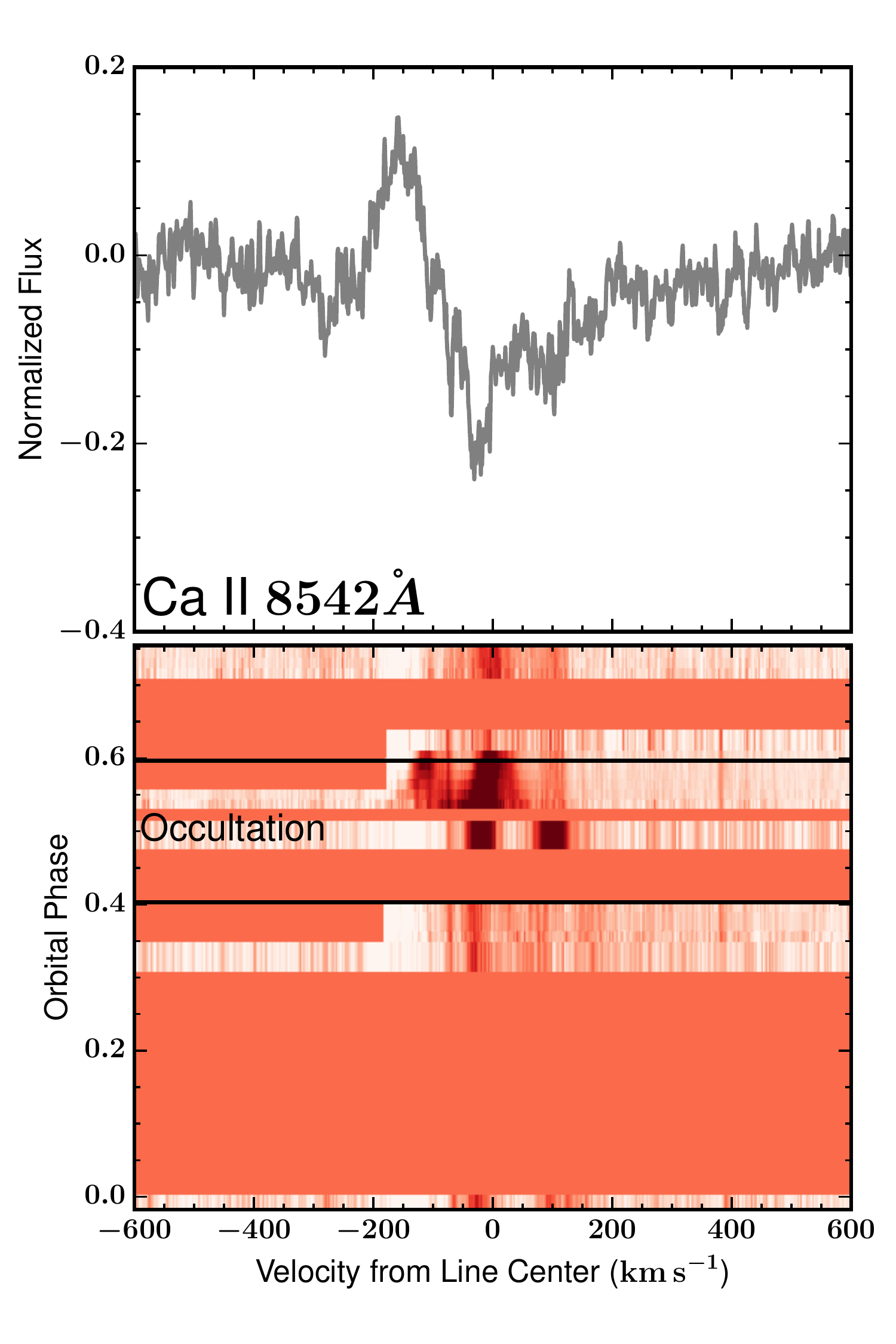} \\
\end{tabular}
\caption{
Spectral signatures of accretion are seen in the spectrum of the primary star, including the wings of the H$\beta$ and H$\alpha$ lines in emission, and the inverted P-Cygni profile of the CaT lines. The out-of-occultation, continuum subtracted, averaged line profiles are shown in the top panels, and the temporal variations are shown on the bottom panels.
\label{fig:accretion_lineprof}}
\end{figure*}

Evidence for active accretion is apparent in the spectra of \thisstar{}. H$\alpha$ is seen strongly in emission during all observations, with an equivalent width of $12\pm2$\,\AA{} measured from the out-of-occultation spectra. The feature is double peaked, and is representative of those seen in longer period $(P>6\,\mathrm{d})$ Algols \citep[e.g.][]{1999ApJS..123..537R,2005ApJ...623..411B}. The wings of the H$\alpha$ feature extend to $\approx 400\,\mathrm{km\,s}^{-1}$, corresponding to an inferred inner disk radius of $\approx 3.7 \,R_\odot$\footnote{The exact extent of the H$\alpha$ wings depends on the spectral normalization and blaze function removal, and is not well constrained.}. In comparison, the isochrone-fitted radius of the $3.24 M_\odot$ accreting star is $\Rone \,R_\odot$ \citep[via fitting of Geneva isochrones,][ as described in section~\ref{sec:system_architecture}]{2012A&A...537A.146E} and, as such, the inner edge of the accretion disk extends close to the stellar surface. The peak emission of both H$\alpha$ and H$\beta$ are located at  $\approx 150\,\mathrm{km\,s}^{-1}$, corresponding to a disk radius of  $\approx 28 \,R_\odot$. The Ca II triplet (CaT) lines exhibit inverted P-Cygni profiles, again indicative of active accretion. The line profiles and phase variations of the accretion features are shown in Figure~\ref{fig:accretion_lineprof}. We see no He features in the accreting star, nor any signatures of wind and outflows in the He I or O I lines. 

The mass-transfer rate onto the accreting star can be estimated via the H$\alpha$ luminosity. By comparing the H$\alpha$ flux against the bolometric luminosity from our SED fit, we measure $L_{\mathrm{H}\alpha} / L_\mathrm{bol,acc} = 0.0023$, implying $L_{\mathrm{H}\alpha} = 0.32_{-0.07}^{+0.15}\,L_\odot$. Assuming complete conversion of potential energy, the mass transfer rate estimated from the H$\alpha$ flux is $1.3\pm0.3\times10^{-8}\,M_\odot \, \mathrm{yr}^{-1}$. For comparison, typical accretion rates measured for Algols in 4 to 13 day period orbits range from $10^{-9}$ to $10^{-5}\,M_\odot\,\mathrm{yr}^{-1}$ \citep[][and references therein]{2016A&A...592A.151V}.  We note that an accurate estimate of the accretion rate should model the H$\alpha$ profile to extract the accretion luminosity, and account for flux from other accretion signatures. Future observations in the UV and more sophisticated modeling of the line profiles will yield more accurate accretion rates for this system. 

\input{paramtable.tex}

\subsection{Gas disk absorption during occultation}
\label{sec:diskspecfeatures}

\begin{figure*} [ht]
\plotone{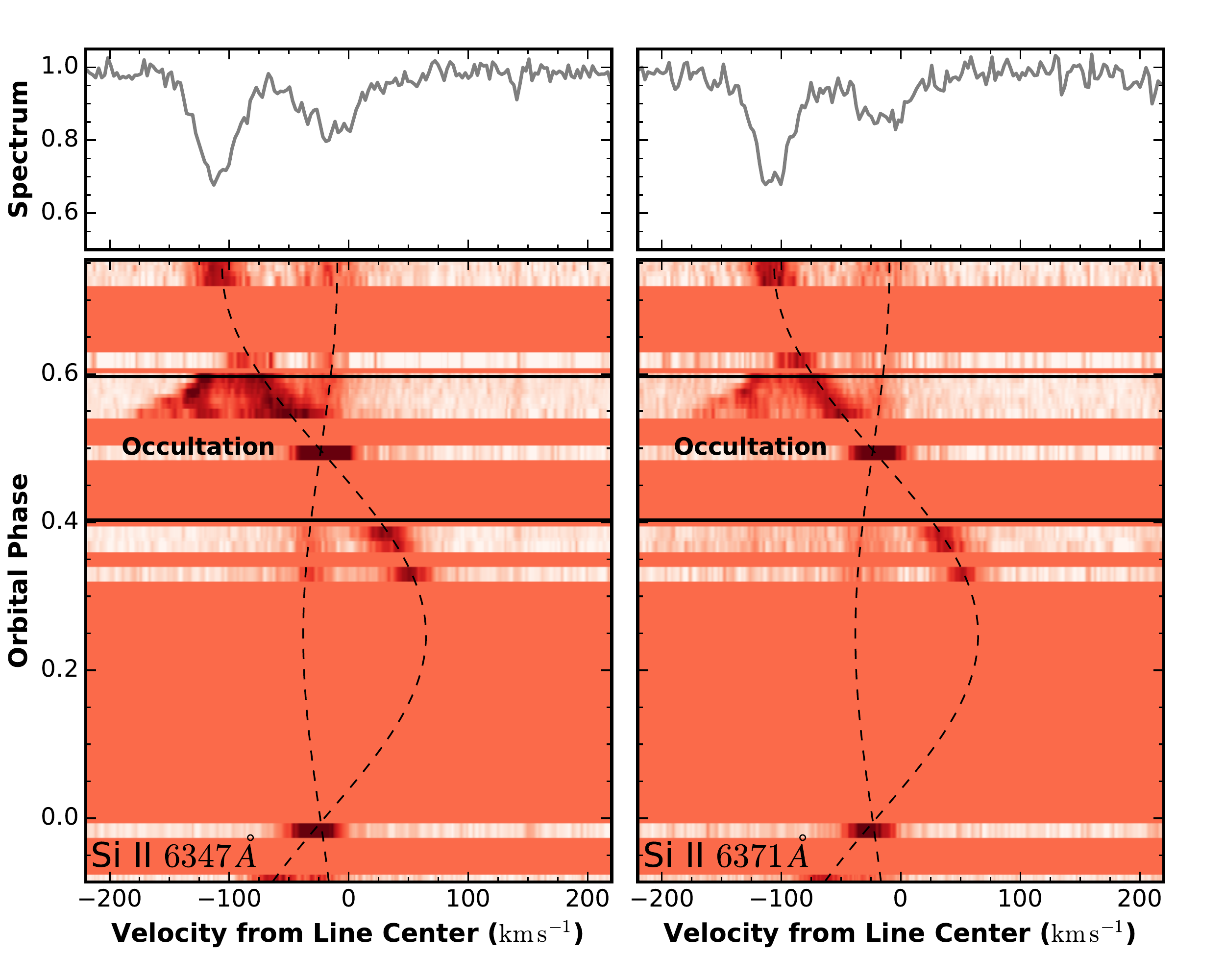}
\caption{
Significant Si II absorption is seen in both stars in the \thisstar{} system. Line strength variations around the Si II lines at 6347\,\AA{} and 6371\,\AA{} are plotted as a function of phase and velocity on the \textbf{bottom panel}, and an illustrative spectral epoch corresponding to the top epoch of the color map shown on the \textbf{top panel}. The Doppler orbit variations, over-plotted by the dashed lines, are seen in both stellar components. Note the appearance of disk absorption features during occultation in both Si II lines, similar in morphology to those seen in the ensemble photospheric lines from the broadening profiles in Figure~\ref{fig:lsd}.
\label{fig:Si6300}}
\end{figure*}  

Occultations like those exhibited by \thisstar{} present opportunities to study the gas material within the accretion disk. During occultation, light from the donor star passes through the accretion disk, and spectroscopic absorption features from the optically thin parts of the gaseous disk are imprinted on the observed spectrum. These absorption lines trace the velocities of disk gas along the line of sight between the donor star and the observer. 

Such occultation disk spectroscopy was obtained during eclipses of $\epsilon$ Aurigae in 1982-1984 \citep{1986PASP...98..389L} and the more recent 2009-2011 event \citep{2012JAVSO..40..729L,2013PASP..125..775G,2014MNRAS.445.2884M,2014AN....335..904S}. The thin dusty and gaseous disk is seen to transit in front of an F0 supergiant, which we view almost edge on (near $90^\circ$ inclination). Disk lines from the optically thin disk `shell' around $\epsilon$ Aurigae were observed at varying widths, depths, and velocities through the two year eclipse. In particular, lines of low excitation potential species were strongly detected, such as Na I, K I, Mg I, H$\alpha$, and H$\beta$. These lines traced the Keplerian velocity of the disk, and helped to constrain the morphology, homogeneity, and potential eccentricity of the disk. 

\begin{figure} [ht]
\plotone{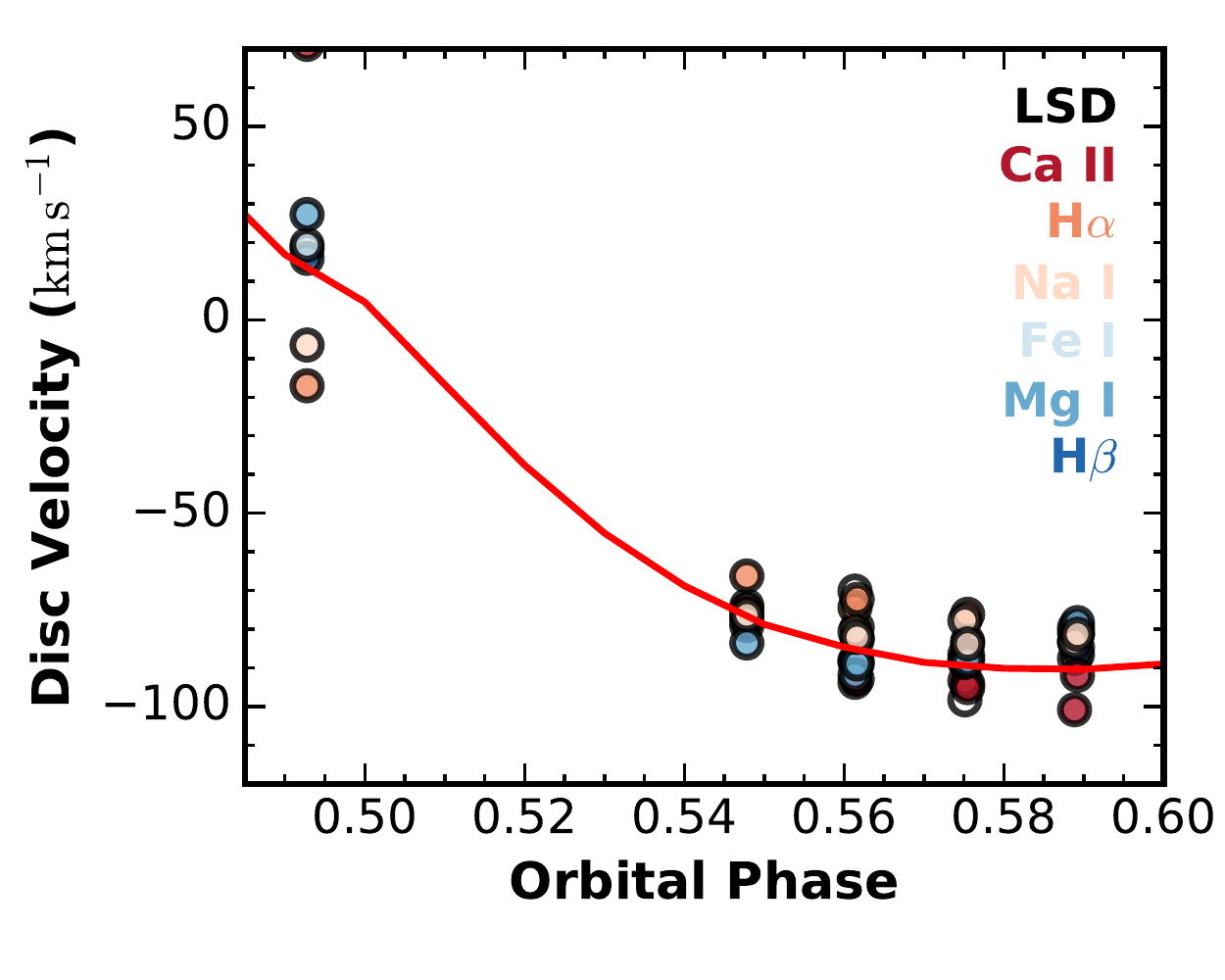}
\caption{
A series of absorption lines from gas in the accretion disk is seen during the occultation. These lines trace out the Keplerian velocity of the disk material being occulted by the donor star. These disk line velocities are used to constrain the disk occultation geometry. Velocities derived from different spectral features are marked by their respective circles. The Keplerian velocity from the best fit disk model is plotted in red.  \label{fig:blobrv}}
\end{figure}

Our spectra obtained during the occultation of \thisstar{} revealed a series of broad absorption features that are offset in velocity from both the donor and accretor.  These features are obvious in the Balmer lines (Figure~\ref{fig:accretion_lineprof}), as well as many neutral and ionised metal lines, such as the Ca II triplet (Figure~\ref{fig:accretion_lineprof}), Si II (Figure~\ref{fig:Si6300}), Fe I, II, Na I D, K I, Mg I. The disk features are also present in the ensemble line profiles derived from the least-squares deconvolution analysis, shown in Figure~\ref{fig:lsd}. The velocities derived from selected unblended lines are shown in Figure~\ref{fig:blobrv}. 

Assuming that the greatest contribution in flux to the disk lines come from the inner-most regions of the disk that actually occult the donor star, we can use these disk line velocities to constrain the occultation geometry and the system architecture (Section~\ref{sec:system_architecture}).

\section{System architecture}
\label{sec:system_architecture}

\begin{figure} [ht]
\plotone{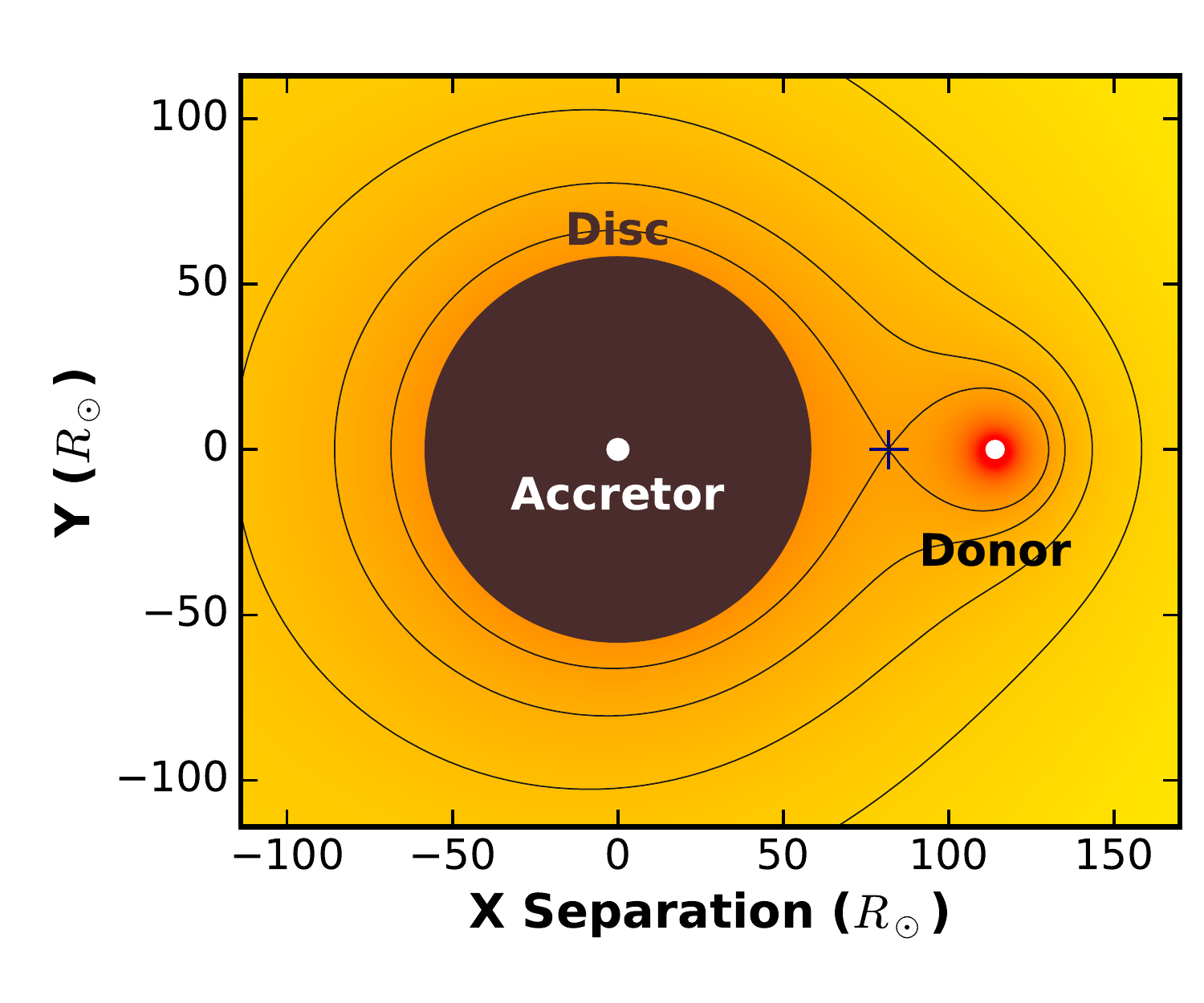}
\caption{
The Roche potential of the binary, with masses of $\Mone\,M_\odot$ and $\Mtwo\,M_\odot$ and separation of $\Rsep \, R_\odot$. Contours of equipotential are marked by the grey lines. The estimated sizes of the two stars, at $\Rone \,R_\odot$ for the accretor and $\Rtwo \,R_\odot$ for the donor, are marked by the white circles. The estimated size of the disk, at $\Rdisk \, R_\odot$ derived from light curve modeling, is marked by the brown circle. The accretion disk is inferred to be nearly filling the Roche lobe of the accreting star. 
\label{fig:roche}}
\end{figure}  

\begin{figure} [ht]
\plotone{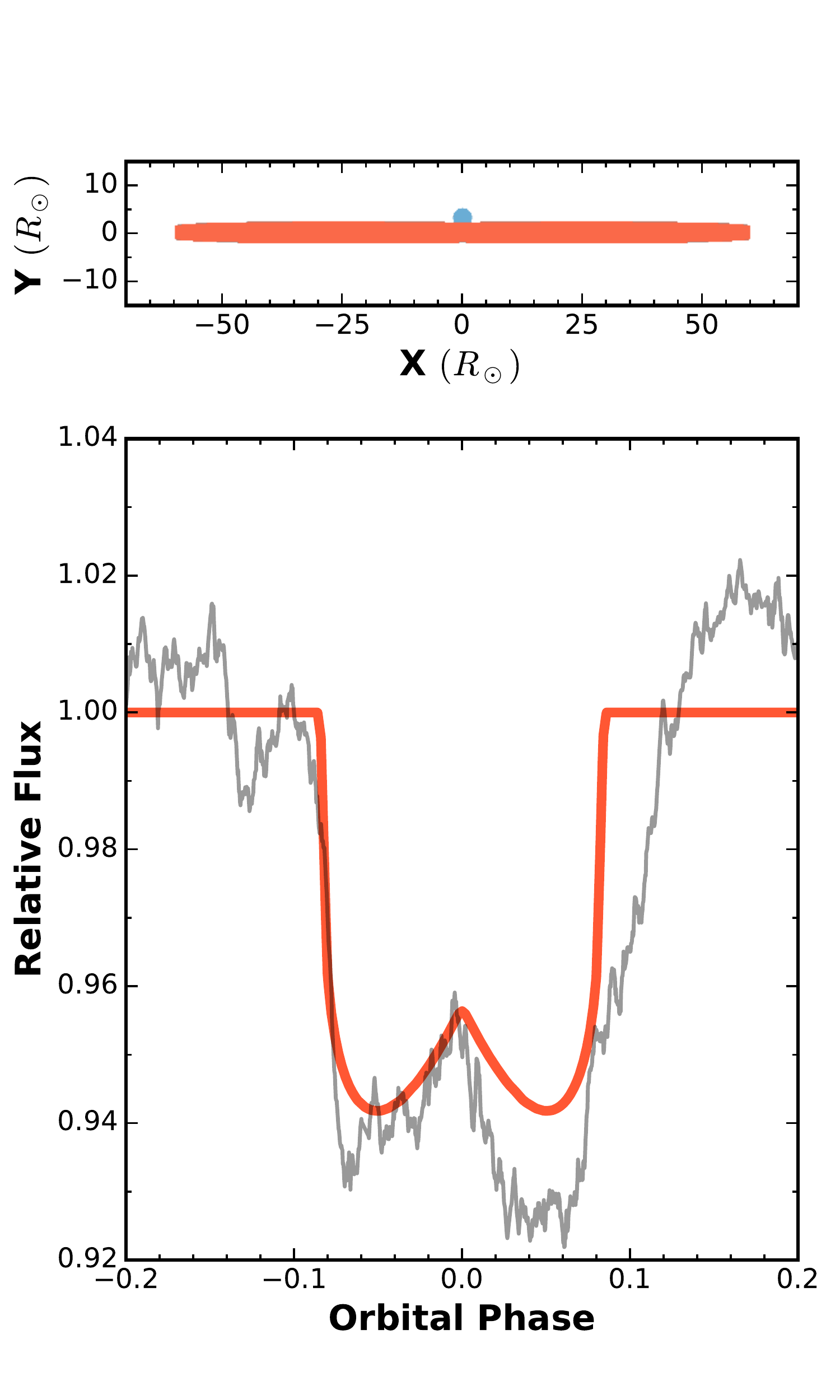}
\caption{
We fit the occultation of a toy disk model to the light curve to derive basic geometric parameters of the system. The best fit configuration involves a disk with radius of $ \Rdisk \,R_\odot$, thickness of $0.63\pm0.21\,R_\odot$ at the inner edge, and exhibiting a $1.63 \pm 0.47 ^\circ$ flare. The entire system inclined to our line of sight by $1.87 \pm 0.58 ^\circ$, occulting a donor star of radius $\Rtwo \,R_\odot$. The eclipse geometry is depicted in the top panel to scale. The model light curve (orange), along with the {\it K2} light curve (grey) are plotted on the bottom panel.
\label{fig:lcmodel}}
\end{figure}  

\begin{figure} [ht]
\plotone{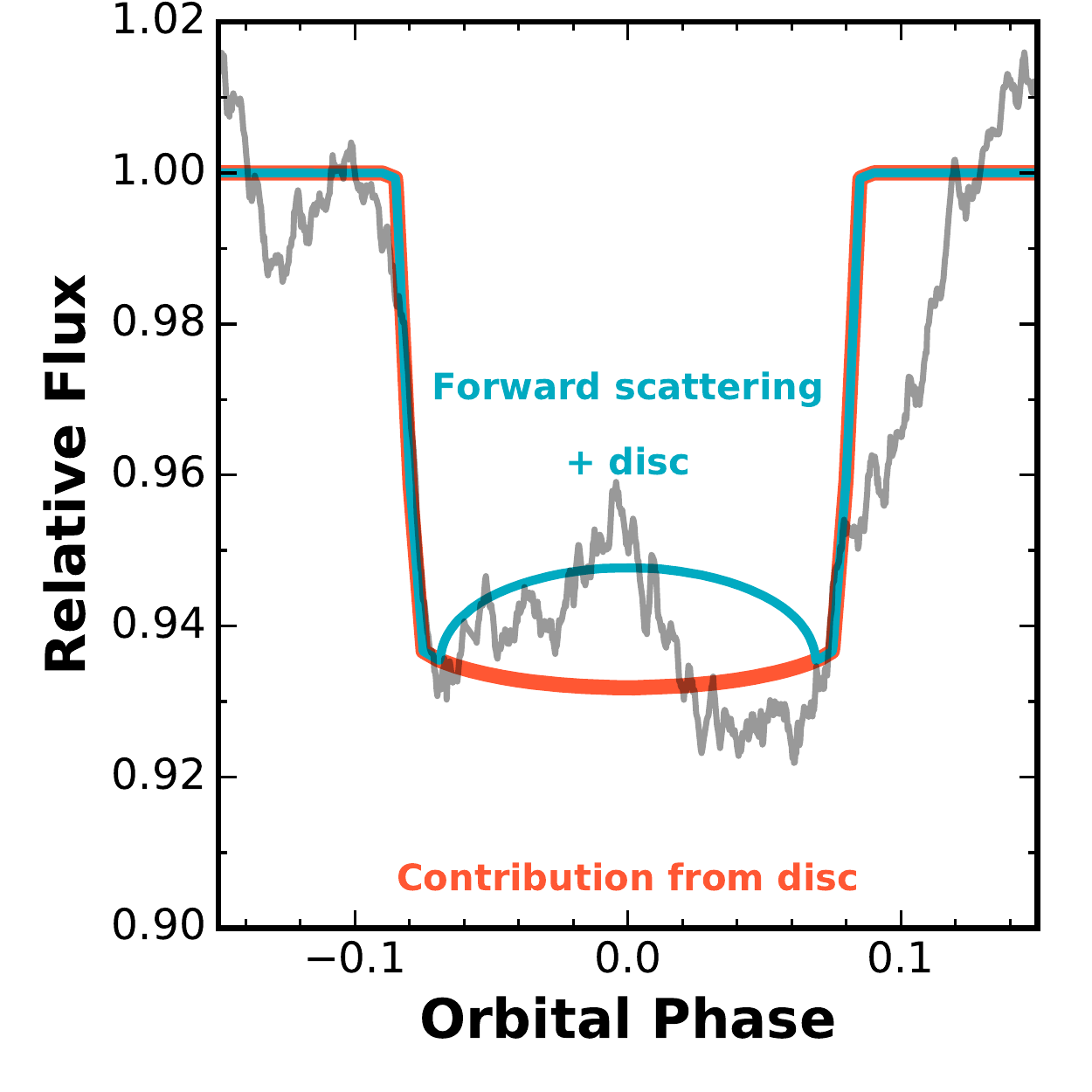}
\caption{
Forward scattering through a dusty disk can also reproduce the central brightening event seen. The red curve shows a flat disk model without forward scattering or flaring of the disk walls. The blue model shows a light curve of the same disk occultation, but accounting for significant forward scattering along the line of sight. 
\label{fig:scatter_lcmodel}}
\end{figure}  

Algols undergoing active accretion are rarely detected at periods as long as 72 days \citep[e.g.][]{1999ApJS..123..537R}. Unravelling the true system architecture is crucial to understanding its evolutionary state. 

From the light curves and radial velocities, we know that the disk around the accreting star is occulting the donor star, the two orbiting each other with a period of 72 days. The dynamical masses indicate the accretor is now $\Mone\,M_\odot$ and the donor is $\Mtwo\,M_\odot$, separated by a distance of $114.0 \pm 3.1\,R_\odot$. Examinations of the spectroscopic lines find that the donor is slightly hotter than the accretor, at 15000\,K and 12000\,K respectively, both contributing approximately equally to the flux of the system. Properties of the accretor can be further estimated by fitting its mass and effective temperature to standard stellar isochrones. Using the Geneva isochrones for high mass stars \citep{2012A&A...537A.146E}, at solar metallicity, and no rotation, the radius and luminosity of the accretor is estimated to be $\Rone \,R_\odot$ and $142_{-33}^{+67}\,L_\odot$. Assuming Stefan-Boltzmann's law and a flux ratio of $F_\mathrm{don} / (F_\mathrm{don} + F_\mathrm{acc}) = 0.52 \pm 0.18$, we calculate the radius of the donor to be $1.79 \pm 0.72\,R_\odot$. We note that the Geneva isochrones are single star evolution tracks, and begin at zero-age main-sequence. The accretor star, however, is a lower mass star that is `rejuvenated' by the mass transfer episode, with different interior structures. Section~\ref{sec:evolution} discusses binary evolution tracks that account for the co-evolution of the two stars. However, the Geneva grid presents a reasonable estimate of the properties of the accretor star, and an interpolation of the grid allows us to estimate uncertainties on the properties of both stars, which is more difficult with the scenario-specific binary evolution tracks. 

Adopting these masses and radii, neither star is close to filling its Roche lobe (Figure~\ref{fig:roche}). The estimated sizes of the Roche lobes are $61.1\pm2.5\,R_\odot$ for the accretor and $27.3\pm 0.9 \,R_\odot$ for the donor \citep[following the approximation in][]{1983ApJ...268..368E}. In fact, a stable disk can only be as large as 80\% of the Roche lobe \citep{1977ApJ...216..822P}, with a maximum radius of $49.4\pm3.2\,R_\odot$. 

We create a toy disk model to simultaneously fit the eclipse light curve and spectroscopic disk velocities. In this model, the eclipse duration constrains the diameter of the disk. The ingress and egress timescales constrain the radius of the donor star, if we assume the disk has a sharp edge and uniform optical depth. The central brightening is modeled by including a flaring to the disk, such that a smaller area of the donor star is covered during mid-eclipse. The velocities of the disk absorption lines seen during occultation are also fitted for simultaneously. At each phase, we take the disk line velocity to be described by the Keplerian orbital velocity of the innermost disk annulus that is occulting the limb of the donor star.

We fit for a disk of radius $R_\mathrm{disk}$ using the following prior. The probability is not penalized if $R_\mathrm{disk} < 49.4$, but follows a Gaussian distribution with $\sigma = 3.2 ~R_\odot$ (see above) if the disk is larger.  We also fit for the donor star radius, $R_\mathrm{don}$, constrained by a Gaussian prior about $1.79 \pm 0.72\,R_\odot$, its inferred radius being taken from spectroscopic analyses. The system is inclined to our line of sight by angle $i$, and the disk exhibits a flare angle of $\beta$, with thickness $h$ at the inner edge, and has uniform optical depth of $\tau$. We make use of all available light curves, and fit for a reference eclipse time $T_c$ and period $P$. We estimated the best-fit values and uncertainties for model parameters with an MCMC exercise. Since the star exhibits photometric jitter only at the 7\,mmag level, the per-point uncertainties are inflated to be the scatter of the out-of-occultation light curve. We find a best fit radius of the donor star of $\Rtwo \,R_\odot$, being occulted by a disk of $R_\mathrm{disk} = \Rdisk \,R_\odot$, with a flaring angle of $1.63 \pm 0.47 ^\circ$ and inclined to our line of sight by $1.87 \pm 0.58 ^\circ$. The best fit light curve model is illustrated in Figure~\ref{fig:lcmodel}, and the best fit parameters summarized in Table~\ref{tab:paramtable}.

Some of the parameters in our model are highly degenerate, and the model is overly simplistic for this system. From the MCMC analysis, the flare angle is highly correlated with the donor star radius and line of sight inclination. The true disk will also not have a sharp edge, and the ingress and egress timescales will be degenerate with the optical depth of the disk and the donor star radius. For future reference, we also note that our adopted priors on disk and donor star radii significantly influence the goodness-of-fit of the light curve. We found that an unconstrained model fits the light curve much better, but tends to adopt a significantly larger disk radius of $\approx 65 \,R_\odot$ and larger donor star of $\approx 15 \,R_\odot$, neither are physical solutions given our known constraints on the system. Even with the Gaussian prior on the disk radius, our derived value of $\Rdisk\,R_\odot$ is in tension with that expected from the Roche lobe constraint at the $1.5\,\sigma$ level. The disk size in our model depends on the eclipse impact factor through a perfectly elliptical projected disk. However, we know from the ingress and egress light curve that the disk is irregular. In this simplistic modeling, we are forcing the disk to be well aligned and symmetric. Asymmetries in the disk can be better explored by future observations that provide better spectroscopic coverage of the disk absorption lines during occultation (Section~\ref{sec:diskspecfeatures}.

It is also possible that the disk contains significant amounts of dust. \thisstar{} exhibits levels of infrared excess consistent with a 1500\,K dusty disk. Similarly, dust $\gtrsim 5$\,$\mu\mathrm{m}$ are thought to be the source of infrared excess in $\epsilon$ Aurigae \citep{2010ApJ...714..549H}. The dust in the disk of $\epsilon$ Aurigae may be partially responsible for the mid-eclipse brightening via forward scattering \citep{2011A&A...532L..12B}. 

Forward scattering can also effectively reproduce the mid-eclipse brightening seen in the occultations of \thisstar{}. During the eclipse, light from the donor star is passing through, or skimming along the surface of the disk. There are three components to the forward scattering pattern: (1) the angular size of the light source as seen by a dust grain; (2) the scattering pattern (i.e., `phase function') vs.~scattering angle for individual dust grains; and (3) the angular size of the dust region as seen from the donor star.  These three angular components would be convolved together. The first of these is just a few degrees due to the small size of the donor. The second component is given approximately from Mie scattering theory by $\sim 20^\circ/s$ where $s$ is the grain size in $\mu$m.  For 5--10 $\mu$m particles, this angle is only $2-4^\circ$.  Finally, the size of the whole dust cloud is essentially $60^\circ$ across, and which dominates the system scattering pattern. As such, at orbital phase $\phi$, the magnitude of the forward scattering effect is roughly proportional to the length of the chord through the disk at each angle $\phi$:
\begin{equation}
B(\phi)  \propto  \sqrt{ \left(R_\mathrm{disk}/a \right)^2 - \sin^2\phi} \, .
\end{equation}

Figure~\ref{fig:scatter_lcmodel} demonstrates the effect of forward scattering on a non-flared disk occultation model. The disk is chosen to be $60\,R_\odot$ wide, inclined to our line of sight at an angle of $1.5^\circ$ with an intrinsic width of $0.5\,R_\odot$. In this demonstration, the magnitude of the scattering effect is arbitrarily normalized to replicate the central brightening seen in the photometry.

\section{System Evolution}
\label{sec:evolution}

We propose that \thisstar{} is a post-Algol binary system. Although there are a number of different configurations for the progenitor binary that could have evolved to match the currently observed system, we believe that the progenitor binary likely consisted of a 3.6 M$ _\odot$ primary (the donor star) and a 2.1 M$ _\odot$ secondary star (accretor) with an orbital period of about 7.05 days. Some of the difficulties in determining the properties of the progenitor binary are related to: (i) uncertainties in the physics describing the amount of mass lost from the binary as it evolves; and, (ii) the magnitude of orbital angular momentum loss due to systemic mass loss, stellar winds, and tidal friction \citep{2000NewAR..44..111E}. It is very difficult to quantify these phenomena but previous theoretical calculations based on reasonable physical assumptions have matched the properties of many observed post-Algol systems \citep[e.g.][and references therein]{1989SSRv...50....1B}. Moreover, the uncertainties can be parameterized and constrained within certain physical limits \citep[e.g.][]{2000NewAR..44..111E}.

\begin{figure*} [ht]
\centering
\includegraphics[width=0.8\linewidth]{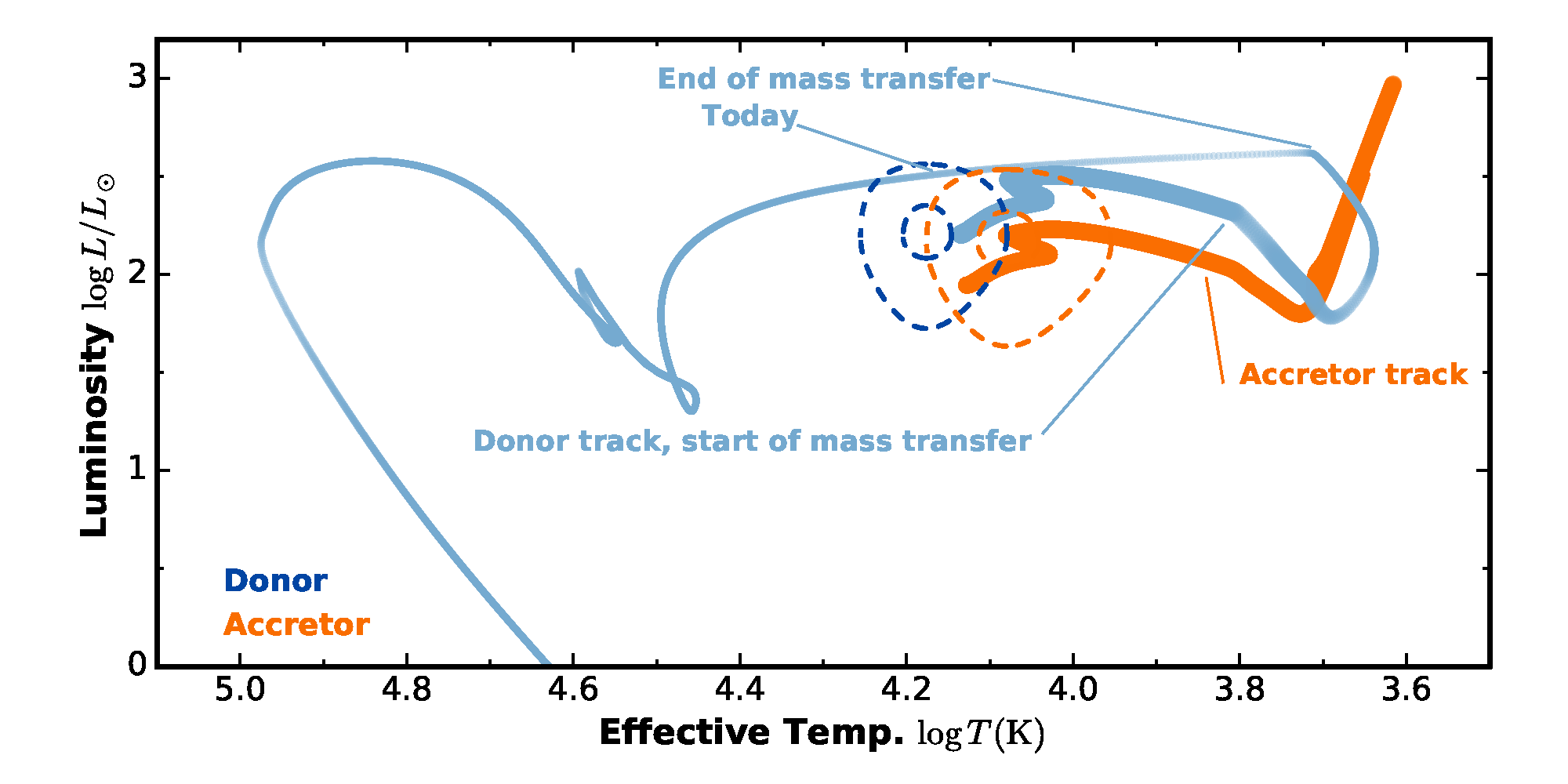} 
\begin{tabular}{cc}
\includegraphics[width=0.4\linewidth]{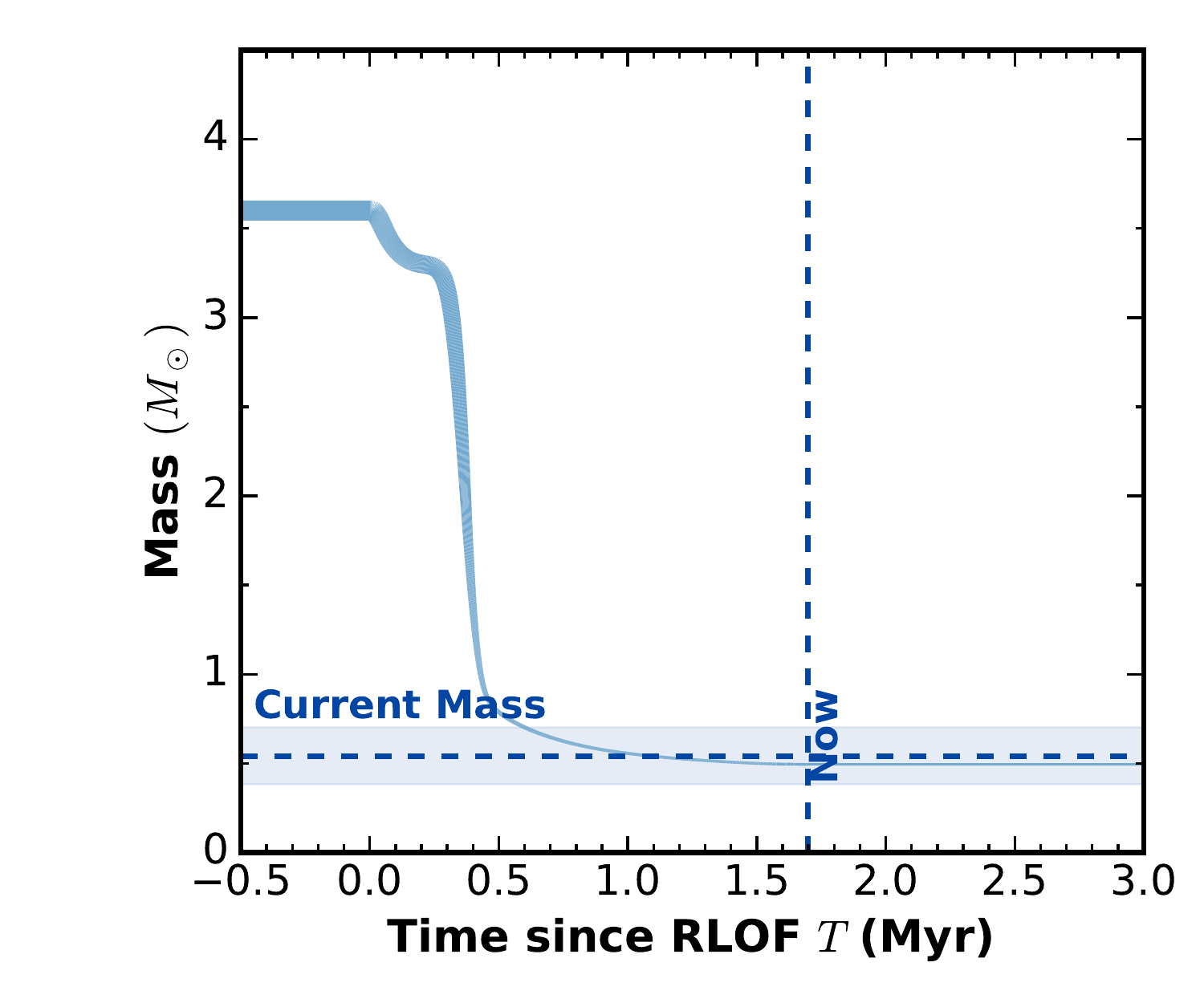} &
\includegraphics[width=0.4\linewidth]{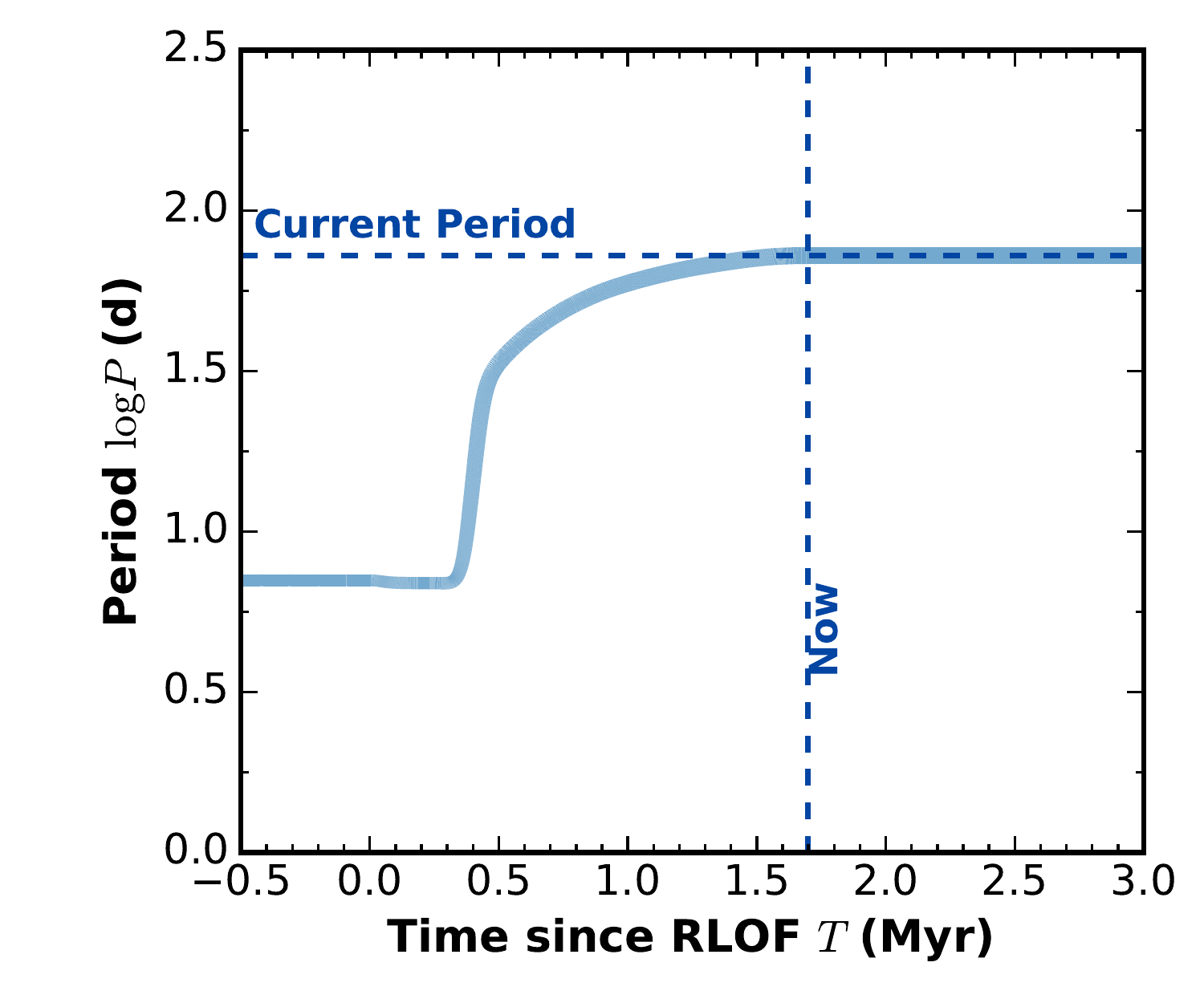} \\
\end{tabular}
\caption{
Self consistent binary evolution tracks for the \thisstar{} system. The evolution of both stars along the HR diagram is plotted on the \textbf{top panel}. The observed luminosity and temperature of the donor (blue) and accretor star (orange) are marked by the $1\sigma$ and $3\sigma$ error ellipses with the dashed lines, the evolution tracks are marked by the solid lines. The original system is composed of a 3.6 M$ _\odot$ donor star and a 2.1 M$ _\odot$ companion in a orbit 7.05 day period orbit. The donor star (blue track) evolved off the main sequence and filled its Roche lobe when it had formed a 0.35 M$ _\odot$ helium core. The mass transfer episode lasts for $\sim$3 Myr. Today, we observe the system about 0.3 Myr after the mass transfer episode ceased and with the donor star contracting along a nearly horizontal track. The accretor's mass has increased to about 3 M$ _\odot$. The width of the evolution tracks corresponds to the mass of the stars, while the density of points mark the time spent in the evolution state. The accretor's future evolution track is plotted in orange. The track begins at its current state as a $3\,M_\odot$ star, and follows the accretor as it evolves through the red-giant branch and the asymptotic-giant branch phases. The \textbf{bottom panels} show the mass and period evolution of the system during the mass transfer episode along the RGB branch. The horizontal axis shows the time since the start of Roche lobe overflow. The width of the line corresponds to the mass of the donor star (as described above). The current masses and periods are labelled by the horizontal lines.  
\label{fig:evolutionary_state}}
\end{figure*}

According to our favored scenario, a 3.6 M$ _\odot$ primary evolves off of the Main Sequence and burns sufficient hydrogen to form a 0.35 M$ _\odot$ helium ash core that is surrounded by a thin hydrogen-burning shell just as it first begins to overflow its Roche lobe. At this point (see Figure~\ref{fig:evolutionary_state}) the luminosity of the donor is about 230 L$ _\odot$ and its radius is $\sim 10$ R$ _\odot$. Mass transfer from the donor star to the 2.1 M$ _\odot$ accreting companion proceeds stably as the donor evolves up the RGB\footnote{Mass transfer that occurs after a helium ash core has formed inside the donor star is known as Case B evolution \citep[see e.g.][]{1985A&A...142..367D}}. Mass-loss rates from the donor can approach 10$^{-5.5}$ M$ _\odot$/yr because of its short nuclear time scale and because of angular momentum carried away by the (systemic) mass-loss from the binary. After the donor has lost about 85{\%} of its mass, it is largely composed of a compact, degenerate helium core with a mass of approximately 0.5 M$ _\odot$, and a very tenuous envelope composed mostly of hydrogen that has a mass of approximately 0.05 M$ _\odot$ and extends over a cross-sectional radius of 25 R$ _\odot$. Continued mass loss cannot be sustained, as this causes this envelope to collapse once a certain threshold in pressure is reached due to the ever decreasing gas densities in the envelope. Thus the donor contracts within its Roche lobe on its thermal (Kelvin-Helmholtz) timescale with the concomitant cessation of mass transfer \citep[see][ and references therein for a detailed discussion]{2004ApJ...616.1124N}. 

Once mass transfer stops, nuclear burning near the surface of the helium core persists for several million years. This causes the surface luminosity to remain high (more than 100 $L_\odot)$ and approximately constant during this phase. As a result, the effective temperature continues to rise as the donor star evolves thermally. We find that the donor star evolves through this post-RGB ``horizontal branch'' phase for approximately 0.35 Myr before attaining a surface temperature and luminosity close to what are inferred for \thisstar. At this juncture the donor has a mass of 0.495\,$M_\odot$ and a luminosity of approximately $300\, L_\odot$. The orbital period of the binary is 72.4 days and the mass of the companion is $3.03\, M_\odot$.

In order to reproduce the observed properties of \thisstar{} a grid of models was computed using the MESA stellar evolution code \citep{2011ApJS..192....3P, 2015ApJS..220...15P}. The evolution of the donor star was followed according to the Roche lobe overflow model. 
The evolution of the accretor was calculated simultaneously, and the computation was stopped after the accretor evolved to become a giant, thereby filling its Roche lobe and leading to a `mass-transfer reversal'.
The chemical composition of the progenitor binary was assumed to be solar (Z$=$0.02) and magnetic braking was assumed to operate in low-mass stars ($\lesssim 1.5 ~M_\odot$) with convective envelopes and radiative cores.

Although highly uncertain, we set the systemic mass-loss and angular momentum-loss parameters such that $\alpha$=0.4, $\beta$=0.3, $\gamma$=0, and $\delta$=0 \citep[for details see][]{2006csxs.book..623T}. Here, $\alpha$ and $\beta$ are fractions of the mass lost by the donor star that get ejected from the binary and carry away the specific angular momentum of the donor star and accreting star, respectively. $\gamma$ and $\delta$ are parameters that are associated with a circumbinary disk and are not used here. Note that these values imply that: (i) mass-transfer is quite non-conservative with an accretion efficiency of 30\% (i.e., with 70\% of the mass being lost via a fast Jeans' ejection); and, (ii) none of the mass lost from the binary forms a circumbinary torus that can extract additional orbital angular momentum.

It is important to note that the track presented in Figure~\ref{fig:evolutionary_state} is only one of several possible tracks that could reproduce the observed properties of the system. This implies that the properties of the progenitor binary (e.g., the initial masses and separation) and the physics of systemic mass and angular-momentum loss (e.g., $\alpha$ and $\beta$) do not have to be fine-tuned in order to match the observations. This increases the probability that the post-Algol hypothesis is correct. 

The donor star in \thisstar{} is currently evolving through the Slowly Pulsating B-star (`SPB') instability strip\footnote{However, the internal structure of this star is radically different from SPB stars that are thought to be close to the main sequence.} and will contract further until substantial helium burning is ignited in its core after $\simeq 3$ Myr has elapsed. During the next $\simeq 50$ Myr the donor star resembles a subwarf B (sdB) star and begins the process of alpha-capture to produce oxygen in the core.  This corresponds to the loop that is seen in Figure~\ref{fig:evolutionary_state} for a temperature of 29,000K and a luminosity of 25\,$L_\odot$.  After an additional $\simeq 35$ Myr, burning at the center of the donor will be complete and the chemical composition at the center is approximately 68\% oxygen and 30\% carbon.  After passing through the sdO (subdwarf O star) region of the HR diagram, the donor reaches its maximum effective temperature (nuclear burning has been quenched) and subsequently enters the degenerate dwarf cooling phase. During this final phase its radius is approximately constant as it evolves to a completely electron degenerate configuration.

We also predict that the accretor will evolve off the main sequence approximately 300 Myr from now and fill its Roche lobe---resulting in a `mass-transfer reversal' episode. The orange track in Figure~\ref{fig:evolutionary_state} illustrates the future evolution of the accretor. At that time, the 3\,$M_\odot$ accretor will have evolved up the AGB and have a 0.54\,$M_\odot$ CO core with a thin helium-burning layer surrounding it. Once mass transfer onto the 0.495\,$M_\odot$ ``donor'' is initiated, it will be dynamically unstable with the ``donor'' spiraling in towards the center of the accretor \citep{1984ApJ...277..355W}.  This common envelope (CE) phase of evolution will last for $\sim$100 to 1000 years, after which the envelope of the accretor will be ejected into the interstellar medium leaving a nascent double-degenerate CO-CO white dwarf binary of nearly equal mass in a tight orbit.  In order to determine the likely orbital period of the double-degenerate binary we follow the approach used by \citet{2017MNRAS.471..948R} (see their Eqn.~10) and set their CE efficiency parameter to 0.5.  
This implies that the remnant binary will have an orbital period of about 1 hour and will be composed of a 0.54\,$M_\odot$ CO white dwarf and a 0.50\,$M_\odot$ CO white dwarf in a nearly circular orbit. The merger of these two white dwarfs as a result of gravitational radiation losses, which will occur within 40 Myrs of the CE phase, may even result in a Type Ia supernova if sub-Chandrasekhar mergers are admissible \citep{2010ApJ...722L.157V}.  Because of the fine-tuning required, it is more likely that the merger product with be a very massive single white dwarf such as inferred for J0317-853 and GD362as (see \citealt{2013ApJ...773..136J} for a discussion of mergers).  If the $0.50\,M_\odot$ white dwarf has a slightly lower mass than our models predict, it is also possible that the merger could lead to the creation of an R CrB star (see, for example, \citealt{2006ApJ...638..454P}). 

Finally, it is important to note that the current-epoch donor star will be a member of the predicted population of subdwarf-B stars in wide binaries around A star accretors while it is in the helium-burning phase of evolution. Our model produces a very natural evolutionary pathway that reproduces the properties of members of this population.  Binary population synthesis by \citet{2003MNRAS.341..669H} predicts a population of donor stars with masses sharply peaked at about $0.5\,M_\odot$, and with orbital periods extending up to $\lesssim 1000$ days, from the first stable Roche lobe overflow channel. The resultant accretors in these systems can have a wide range of spectral types based on their masses and degree of nuclear evolution. We could reasonably expect that most unevolved accretors would fall within the F to B V spectral classes; the evolved systems could have much cooler temperatures but are less likely to be observed because of their relatively short lifetimes. They note that most of these systems are probably beyond the detection thresholds of typical radial velocity surveys \citep[e.g.][]{2001MNRAS.326.1391M} due to the small orbital semi-amplitudes of the luminous accretor stars.

\subsection{Why is the disk still around?}
\label{sec:disk_puzzle}

In the scenario outlined above, active mass transfer ended $\sim 0.3$\,Myr ago when the donor star evolved off the RGB, and under-filled its Roche lobe. The occultation we observe, however, indicates that a substantial disk of size $\Rdisk\,R_\odot$ still remains around the accretor. We explore a few simple arguments for the lifetimes of gas and dusty disks below, to show that without the disk being actively fed, it is difficult for the disk to remain in its current state. 

To estimate the lifetime of a gaseous disk, we adopt the \citet{1973A&A....24..337S} $\alpha$-disk model for an approximation of the disk filling timescale $\tau_\mathrm{fill}$:
\begin{align}
  \tau_\mathrm{fill}  &= \frac{M_\mathrm{disk}} {\dot{M}}  \nonumber{} \\
& \simeq 240 \left(\frac{\alpha}{0.01}\right)^{-4/5} \left( \frac{\dot{M}}{10^{-8} M_\odot / \mathrm{yr}} \right)^{-3/10} \left(  \frac{r_\mathrm{max}}{60\, R_\odot} \right)^{5/4} \,\mathrm{yr}\,, 
\end{align}
where $\dot{M}$ is the accretion rate, and $r_\mathrm{max}$ is the outer radius of the disk, and the parameter $\alpha$ specifies the efficiency of viscous angular momentum transport in the disk. Adopting our H$\alpha$ derived accretion rate of $10^{-8}\,M_\odot\,\mathrm{yr}^{-1}$, and an $\alpha$ parameter of 0.01, the disk should have dissipated in just 200 years. In fact, while our accretion rate may be somewhat underestimated, it needs to be smaller by 10 orders of magnitude to achieve a disk lifetime of 0.2 Myr. Alternatively, $\alpha$ would have to be $2 \times 10^{-6}$ for the disk to last for 0.25 Myr.  We also note that for the disk to maintain its current accretion rate of $1.3 \times 10^{-8}\,M_\odot\,\mathrm{yr}^{-1}$ over 0.2\,Myr, the disk needs to be more massive than $0.003 \, M_\odot$. However, given that the currently inferred accretion rate is similar in magnitude to the theoretically expected mass transfer rate during the Roche-lobe-filling phase, it would be difficult to build up such a massive disk. 

\begin{figure} [ht]
\centering
\includegraphics[width=1\linewidth]{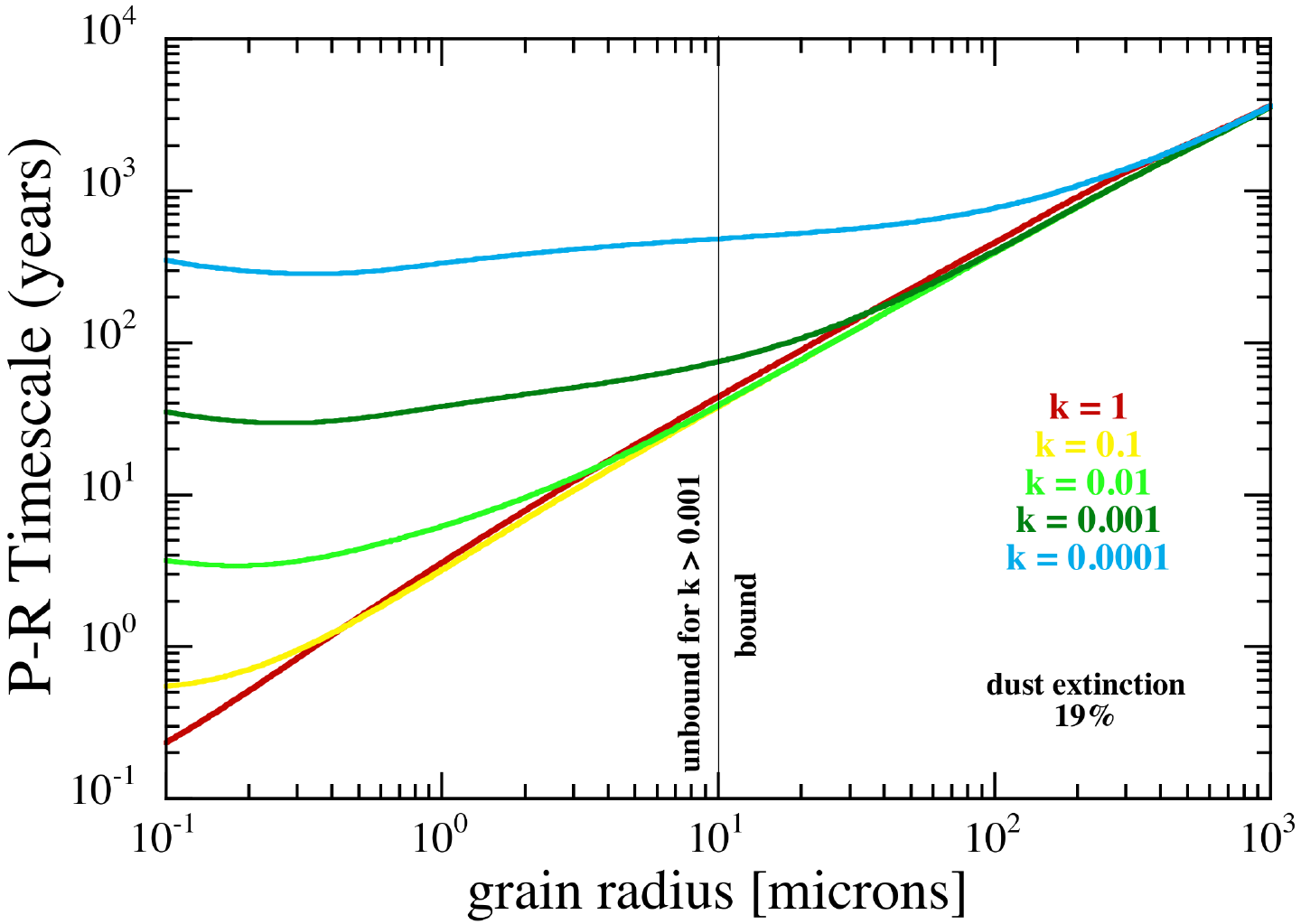}
\caption{A dusty disk is subjected to Poynting-Robertson orbital decay if the dust is relatively optically thin. We plot the orbital decay timescale for dust particles of various grain sizes and imaginary refraction indices $k$ when placed in an $49\,R_\odot$ orbit around the accretor.  These P-R timescales have been computed using only the dust {\it absorption} cross sections, thereby providing an upper limit to the orbital decay timescales.  Dust particles smaller than $100\,\mu$m experience decay on time scales no more than $\sim 10^3$ years, and are not sufficient in sustaining the disk seen around \thisstar{} today. The vertical line at 10 $\mu$m indicates the dividing line where, for smaller particles, radiation pressure forces will directly unbind them from the system.
\label{fig:dust_timescales}}
\end{figure}

We can also explore the possibility that the disk contains significant dust. Dust particles in the disk are subjected to the Poynting-Robertson drag, by which dust grains lose orbital angular momentum. This results from the fact that in the rest frame of the dust there is a small component of the momentum flux of the photons that is in the direction opposed to the orbital motion.  Figure~\ref{fig:dust_timescales} shows the expected orbital decay timescales from Poynting-Robertson drag on dust particles in the disk at $49\,R_\odot$, for dust having different imaginary indices of refraction $k$. As can be seen from Figure~\ref{fig:dust_timescales}, such dust particles have orbital lifetimes of no more than $\sim$1000 years, insufficient to sustain the disk in its current state.  Furthermore, most dust particles smaller than $\sim$10 $\mu$m have ratios of radiation-pressure forces to gravity $\gtrsim 0.5$ and therefore they become unbound from the system on a dynamical timescale.

\section{Summary}

\thisstar{} is a post-Algol binary with a B7 post-RGB donor star that, until $\sim 0.3$\,Myr ago, was transferring mass to its A0 companion. The donor star, now pushed outward to an orbital separation of $\Rsep \,R_\odot$, is occulted by the remnant accretion disk around the accretor once every 72 days. The occultations were observed during Campaign 11 of the {\it K2} mission, and subsequently identified in pre-discovery light curves from the ASAS and ASAS-SN surveys. The dynamical masses of the system were measured to be $\Mone\,M_\odot$ and $\Mtwo\,M_\odot$. The strengths of temperature sensitive lines yielded a spectral type estimate of A0 for the accretor and B7 for the donor star, with the two stars exhibiting approximately equal luminosities. We estimate, via isochrone fitting and light curve modeling that the radii of the two stars are $\Rone\,R_\odot$ and $\Rtwo\,R_\odot$.

We coordinated a campaign of multi-band photometric and spectroscopic observations over the occultation event centered on September 2017. Our eclipse light curves agreed well with those observed by {\it K2} and pre-discovery surveys. The eclipse was found to be colorless to within detection limits. We obtained a series of spectroscopic observations over the second half of the eclipse, and detected a series of absorption lines from the disk around the accretor, reminiscent of those seen during the eclipse of $\epsilon$ Aurigae \citep[e.g.][]{2011A&A...530A.146C,2014AN....335..904S}. We used a toy disk model to simultaneously fit for the photometric and spectroscopic eclipses, finding a disk $\Rdisk \,R_\odot$ in radius. The central brightening seen during eclipse is explained by including a flaring geometry to the disk, such that a smaller area of the donor star is covered during mid-eclipse, or by a dusty disk inducing significant forward scattering along the line of sight \citep[similar to the disk model of $\epsilon$ Aurigae from][]{2011A&A...532L..12B}. 

Our interpretation of the system cannot account for the persistence of the accretion disk. We expect the donor star to have begun contraction $\sim0.3$\,Myr ago, terminating mass transfer, while the disk is nominally expected to survive for only hundreds of years without active feeding, regardless of its dust to gas composition. We suggest future observations could search for signatures of active mass transfer in the system. Full spectroscopic coverage of the orbital phase can reconstruct a Doppler tomographic image of the inner disk \citep[e.g.][]{1988MNRAS.235..269M}, and help map out the ongoing accretion mechanics. Sporadic hot spots in the disk are also signs of mass transfer, and may be identified by frequent monitoring of the system at all phases. In particular, higher signal-to-noise spectra during the eclipse, covering ingress and egress, can help us constrain the optical depths at the edges of the disk, and allow us to resolve degeneracies that plague the light curve modeling. The disk of $\epsilon$ Aurigae is thought to consist of large dust grains \citep[e.g.][]{2010ApJ...714..549H}, which may help extend the disk lifetime. Similar mid- and far-infrared observations of \thisstar{} may constrain the disk gas to dust ratio, and help better understand why the disk is still present. Continuous spectroscopic and spectro-polarimetric monitoring of the system over the rotation period of the inner disk and the stars may reveal connections between accretion and the chemical peculiarity of the stars. Ap/Bp stars are known to exhibit strong magnetic fields, spots, and inhomogeneities in the abundances across their surfaces \citep[e.g.][]{2004A&A...414..613K,2010A&A...513A..13K}. We can search for links between the inner disk structure and the potential spot distribution in the stars of \thisstar{}.

\acknowledgements  

Work by G.Z. is provided by NASA through Hubble Fellowship grant HST-HF2-51402.001-A awarded by the Space Telescope Science Institute, which is operated by the Association of Universities for Research in Astronomy, Inc., for NASA, under contract NAS 5-26555.
L.N. thanks the Natural Sciences and Engineering Research Council (Canada) for financial support provided through a Discovery grant.
C.I.J. gratefully acknowledges support from the Clay Fellowship, administered by the Smithsonian Astrophysical Observatory.
M.H.K would like to acknowledge Allan Schmitt for his LcTools software.
Work by C.H. is supported by the Juan Carlos Torres Fellowship.
Work by A.V is performed in part under contract with the California Institute of Technology/Jet Propulsion Laboratory funded by NASA through the Sagan Fellowship Program executed by the NASA Exoplanet Science Institute. 
This work makes use of the Smithsonian Institution High Performance Cluster (SI/HPC) 
We also thank Calcul Qu\'ebec, the Canada Foundation for Innovation (CFI), NanoQu\'{e}bec, RMGA, and the Fonds de recherche du Qu\'{e}bec - Nature et technologies (FRQNT) for computational facilites.  We would also like to thank J. Cannizzo for insightful discussions.
This paper includes data taken at The McDonald Observatory of The University of Texas at Austin. 

Facilities: 
\facility{ASAS, ASASSN, Kepler, FLWO:1.5m (TRES), Smith (Tull), APF, HAO, PEST}

\bibliographystyle{apj}
\bibliography{mybib}

\end{document}

%% file: literature_params.tex
\begin{deluxetable*}{lrr}
\tablewidth{0pc}
\tabletypesize{\scriptsize}
\tablecaption{
    Photometric and astrometric parameters
    \label{tab:litparams}
}
\tablehead{
    \multicolumn{1}{c}{~~~~~~~~Parameter~~~~~~~~}   &
    \multicolumn{1}{c}{Value} &
    \multicolumn{1}{c}{Source} \\
}
\startdata
~~~Bonner Durchmusterung (BD)                    \dotfill    & BD-22 4376\\
~~~EPIC ID                                       \dotfill    & 225300403 & \\
~~~2MASS ID                                      \dotfill    & J17362552-2246179 & \\
~~~Tycho ID                                      \dotfill    & TYC 6827-410-1\\

~~~RA (J2000)    \dotfill    & 17 36 25.5215 & \\
~~~DEC (J2000)   \dotfill    & -22 46 17.904 & \\
~~~NUV (mag) \dotfill & $15.742\pm0.017$ & GALEX {\citep{2011Ap&SS.335..161B}} \\
~~~$BJ$ (mag) \dotfill & $11.376\pm0.100$ & ASCC2.5V3  {\citep{2009yCat.1280....0K}} \\
~~~$BJ$ (mag) \dotfill & $11.409\pm0.017$ & APASS DR9 {\citep{2016yCat.2336....0H}} \\
~~~$VJ$ (mag) \dotfill & $10.935\pm0.112$ & ASCC2.5V3  {\citep{2009yCat.1280....0K}} \\
~~~$VJ$ (mag) \dotfill & $10.802\pm0.030$ & APASS DR9 {\citep{2016yCat.2336....0H}} \\
~~~$g'$ (mag) \dotfill & $11.093\pm0.017$ & APASS DR9 {\citep{2016yCat.2336....0H}} \\
~~~$r'$ (mag) \dotfill & $10.616\pm0.040$ & APASS DR9 {\citep{2016yCat.2336....0H}} \\
~~~$i'$ (mag) \dotfill & $10.360\pm0.016$ & APASS DR9 {\citep{2016yCat.2336....0H}}\\
~~~$J$ (mag) \dotfill & $9.298\pm0.024$ & 2MASS {\citep{2003tmc..book.....C}}  \\
~~~$H$ (mag) \dotfill & $9.044\pm0.028$ & 2MASS {\citep{2003tmc..book.....C}} \\
~~~$K$ (mag) \dotfill & $8.892\pm0.025$ & 2MASS {\citep{2003tmc..book.....C}} \\
~~~$W1$ (mag) \dotfill & $8.560\pm0.023$ & WISE {\citep{2012yCat.2311....0C}} \\
~~~$W2$ (mag) \dotfill & $8.395\pm0.022$ & WISE {\citep{2012yCat.2311....0C}} \\
~~~$W3$ (mag) \dotfill & $7.792\pm0.023$ & WISE {\citep{2012yCat.2311....0C}} \\
~~~$W4$ (mag) \dotfill & $7.44\pm0.21$ & WISE {\citep{2012yCat.2311....0C}} \\
~~~Line of sight reddening $E(B-V)$ (mag) \dotfill & $0.866 \pm  0.024$ & IRAS \citep{1998ApJ...500..525S} \\
~~~Fitted system reddening $E(B-V)$ (mag) \dotfill & 1.07 & SED fitting \\

\enddata
\end{deluxetable*}

%% file: paramtable.tex
\begin{deluxetable*}{lrr}
\tablewidth{0pc}
\tabletypesize{\scriptsize}
\tablecaption{
    System parameters
    \label{tab:paramtable}
}
\tablehead{
    \multicolumn{1}{c}{~~~~~~~~Parameter~~~~~~~~}   &
    \multicolumn{1}{c}{Value} &
    \multicolumn{1}{c}{Source} \\
}
\startdata
\sidehead{\Lc{} parameters}
~~~Period $P$ (days)                                    \dotfill    & $72.416 \pm 0.016$ & Light curve modeling \\
~~~Occultation centroid $T_c$ (${\rm BJD}$)             \dotfill    & $2457713.18 \pm 0.18$ & Light curve modeling \\

\sidehead{RV parameters}
~~~$K\mathrm{rv} _1$ (\kms)              \dotfill    & \Kone{} & Radial velocity modeling\\
~~~$K\mathrm{rv} _2$ (\kms)              \dotfill    & \Ktwo{} & Radial velocity modeling\\
~~~$e\cos\omega$                         \dotfill    & $-0.0127\pm 0.0097$ & Radial velocity modeling\\
~~~$e\sin\omega$                         \dotfill    & $-0.003 \pm 0.017$ & Radial velocity modeling\\
~~~Systemic RV (\kms)                    \dotfill    & $-22.7 \pm 1.9$ & Radial velocity modeling \\

\sidehead{Derived parameters}
~~~$M_\mathrm{acc}\,(M_\odot)$       \dotfill    & \Mone{} & Radial velocity modeling \\
~~~$R_\mathrm{acc}\,(R_\odot)$       \dotfill    & \Rone{} & Spectral + Geneva Isochrone fitting {\citep{2012A&A...537A.146E}} \\    
~~~$L_\mathrm{acc}\,(L_\odot)$       \dotfill    & $142_{-33}^{+67}$ & Spectral + Geneva Isochrone fitting {\citep{2012A&A...537A.146E}} \\
~~~$M_\mathrm{don}\,(M_\odot)$       \dotfill    & \Mtwo{} & Radial velocity modeling \\
~~~$R_\mathrm{don}\,(R_\odot)$       \dotfill    & $1.79 \pm 0.72$ & Spectral + Geneva Isochrone fitting {\citep{2012A&A...537A.146E}} \\ 
~~~$R_\mathrm{don}\,(R_\odot)$       \dotfill    & $2.01 \pm 0.52$ & Disk modeling + Gaussian prior from isochrone fitting \\     
~~~$L_\mathrm{don}/(L_\mathrm{don}+L_\mathrm{acc})$       \dotfill    & $0.52 \pm 0.18$ & Spectral fitting \\    
~~~Distance (pc)                         \dotfill   &     $770 \pm 140$ & Isochrone fitting \\
~~~$a\,(R_\odot)$               \dotfill    & \Rsep{} & Radial velocity modeling \\
 ~~~$e$                    \dotfill    & $0.021 \pm 0.010$ & Radial velocity modeling  \\

\sidehead{Derived disk parameters}
~~~$R_\mathrm{disk}\,(R_\odot)$                     \dotfill    & \Rdisk{} & Light curve modeling and Roche-lobe constraint \\ 
~~~$R_\mathrm{disk}\,(R_\odot)$                     \dotfill    & $49.2 \pm 3.2$ & From Roche-lobe constraint only \\ 
~~~Line of sight inclination $inc\,(^\circ)$        \dotfill    & $1.43 \pm 0.56$ & Light curve modeling \\
~~~Disk flare angle $\beta \, (^\circ)$             \dotfill  & $0.70 \pm 0.25$ & Light curve modeling \\
~~~Disk thickness $h\,(R_\odot)$                       \dotfill  & $0.63\pm0.21$ & Light curve modeling \\ 
~~~Disk optical depth $\tau$            \dotfill  & $0.19\pm0.13$ & Light curve modeling \\

\enddata
\end{deluxetable*}

%% file: paper_revision_nobold.bbl
\begin{thebibliography}{95}
\expandafter\ifx\csname natexlab\endcsname\relax\def\natexlab#1{#1}\fi

\bibitem[{{Bassett}(1978)}]{1978Obs....98..122B}
{Bassett}, E.~E. 1978, The Observatory, 98, 122

\bibitem[{{Batten}(1989{\natexlab{a}})}]{1989SSRv...50.....B}
{Batten}, A.~H. 1989{\natexlab{a}}, \ssr, 50

\bibitem[{{Batten}(1989{\natexlab{b}})}]{1989SSRv...50....1B}
---. 1989{\natexlab{b}}, \ssr, 50, 1

\bibitem[{{Bianchi} {et~al.}(2011){Bianchi}, {Herald}, {Efremova}, {Girardi},
  {Zabot}, {Marigo}, {Conti}, \& {Shiao}}]{2011Ap&SS.335..161B}
{Bianchi}, L., {Herald}, J., {Efremova}, B., {et~al.} 2011, \apss, 335, 161

\bibitem[{{Buchhave} {et~al.}(2010){Buchhave}, {Bakos}, {Hartman}, {Torres},
  {Kov{\'a}cs}, {Latham}, {Noyes}, {Esquerdo}, {Everett}, {Howard}, {Marcy},
  {Fischer}, {Johnson}, {Andersen}, {F{\H u}r{\'e}sz}, {Perumpilly},
  {Sasselov}, {Stefanik}, {B{\'e}ky}, {L{\'a}z{\'a}r}, {Papp}, \&
  {S{\'a}ri}}]{2010ApJ...720.1118B}
{Buchhave}, L.~A., {Bakos}, G.~{\'A}., {Hartman}, J.~D., {et~al.} 2010, \apj,
  720, 1118

\bibitem[{{Budaj}(2011)}]{2011A&A...532L..12B}
{Budaj}, J. 2011, \aap, 532, L12

\bibitem[{{Budaj} {et~al.}(2005){Budaj}, {Richards}, \&
  {Miller}}]{2005ApJ...623..411B}
{Budaj}, J., {Richards}, M.~T., \& {Miller}, B. 2005, \apj, 623, 411

\bibitem[{{Burt} {et~al.}(2015){Burt}, {Holden}, {Hanson}, {Laughlin}, {Vogt},
  {Butler}, {Keiser}, \& {Deich}}]{Burt2015}
{Burt}, J., {Holden}, B., {Hanson}, R., {et~al.} 2015, Journal of Astronomical
  Telescopes, Instruments, and Systems, 1, 044003

\bibitem[{{Carroll} {et~al.}(1991){Carroll}, {Guinan}, {McCook}, \&
  {Donahue}}]{1991ApJ...367..278C}
{Carroll}, S.~M., {Guinan}, E.~F., {McCook}, G.~P., \& {Donahue}, R.~A. 1991,
  \apj, 367, 278

\bibitem[{{Castelli} \& {Kurucz}(2004)}]{2004astro.ph..5087C}
{Castelli}, F., \& {Kurucz}, R.~L. 2004, ArXiv Astrophysics e-prints

\bibitem[{{Chadima} {et~al.}(2011){Chadima}, {Harmanec}, {Bennett},
  {Kloppenborg}, {Stencel}, {Yang}, {Bo{\v z}i{\'c}}, {{\v S}lechta},
  {Kotkov{\'a}}, {Wolf}, {{\v S}koda}, {Votruba}, {Hopkins}, {Buil}, \&
  {Sudar}}]{2011A&A...530A.146C}
{Chadima}, P., {Harmanec}, P., {Bennett}, P.~D., {et~al.} 2011, \aap, 530, A146

\bibitem[{{Cutri} \& {et al.}(2012)}]{2012yCat.2311....0C}
{Cutri}, R.~M., \& {et al.} 2012, VizieR Online Data Catalog, 2311

\bibitem[{{Cutri} {et~al.}(2003){Cutri}, {Skrutskie}, {van Dyk}, {Beichman},
  {Carpenter}, {Chester}, {Cambresy}, {Evans}, {Fowler}, {Gizis}, {Howard},
  {Huchra}, {Jarrett}, {Kopan}, {Kirkpatrick}, {Light}, {Marsh}, {McCallon},
  {Schneider}, {Stiening}, {Sykes}, {Weinberg}, {Wheaton}, {Wheelock}, \&
  {Zacarias}}]{2003tmc..book.....C}
{Cutri}, R.~M., {Skrutskie}, M.~F., {van Dyk}, S., {et~al.} 2003, {2MASS All
  Sky Catalog of point sources.}

\bibitem[{{De Greve} {et~al.}(1985){De Greve}, {Packet}, \& {de
  Landtsheer}}]{1985A&A...142..367D}
{De Greve}, J.~P., {Packet}, W., \& {de Landtsheer}, A.~C. 1985, \aap, 142, 367

\bibitem[{{De Loore} \& {van Rensbergen}(2005)}]{2005Ap&SS.296..353D}
{De Loore}, C., \& {van Rensbergen}, W. 2005, \apss, 296, 353

\bibitem[{{Dervi{\c s}o{\v g}lu} {et~al.}(2010){Dervi{\c s}o{\v g}lu}, {Tout},
  \& {Ibano{\v g}lu}}]{2010MNRAS.406.1071D}
{Dervi{\c s}o{\v g}lu}, A., {Tout}, C.~A., \& {Ibano{\v g}lu}, C. 2010, \mnras,
  406, 1071

\bibitem[{{Donati} {et~al.}(1997){Donati}, {Semel}, {Carter}, {Rees}, \&
  {Collier Cameron}}]{1997MNRAS.291..658D}
{Donati}, J.-F., {Semel}, M., {Carter}, B.~D., {Rees}, D.~E., \& {Collier
  Cameron}, A. 1997, \mnras, 291, 658

\bibitem[{{Dong} {et~al.}(2014){Dong}, {Katz}, {Prieto}, {Udalski},
  {Kozlowski}, {Street}, {Bramich}, {Tsapras}, {Hundertmark}, {Snodgrass},
  {Horne}, {Dominik}, \& {Figuera Jaimes}}]{2014ApJ...788...41D}
{Dong}, S., {Katz}, B., {Prieto}, J.~L., {et~al.} 2014, \apj, 788, 41

\bibitem[{{Duch{\^e}ne} \& {Kraus}(2013)}]{2013ARA&A..51..269D}
{Duch{\^e}ne}, G., \& {Kraus}, A. 2013, \araa, 51, 269

\bibitem[{{Eggleton}(1983)}]{1983ApJ...268..368E}
{Eggleton}, P.~P. 1983, \apj, 268, 368

\bibitem[{{Eggleton}(2000)}]{2000NewAR..44..111E}
---. 2000, New Astronomy Reviews, 44, 111

\bibitem[{{Eggleton} \& {Kiseleva-Eggleton}(2002)}]{2002ApJ...575..461E}
{Eggleton}, P.~P., \& {Kiseleva-Eggleton}, L. 2002, \apj, 575, 461

\bibitem[{{Ekstr{\"o}m} {et~al.}(2012){Ekstr{\"o}m}, {Georgy}, {Eggenberger},
  {Meynet}, {Mowlavi}, {Wyttenbach}, {Granada}, {Decressin}, {Hirschi},
  {Frischknecht}, {Charbonnel}, \& {Maeder}}]{2012A&A...537A.146E}
{Ekstr{\"o}m}, S., {Georgy}, C., {Eggenberger}, P., {et~al.} 2012, \aap, 537,
  A146

\bibitem[{{Foreman-Mackey} {et~al.}(2013){Foreman-Mackey}, {Hogg}, {Lang}, \&
  {Goodman}}]{2013PASP..125..306F}
{Foreman-Mackey}, D., {Hogg}, D.~W., {Lang}, D., \& {Goodman}, J. 2013, \pasp,
  125, 306

\bibitem[{{Gray}(2005)}]{2005oasp.book.....G}
{Gray}, D.~F. 2005, {The Observation and Analysis of Stellar Photospheres}

\bibitem[{{Gray} \& {Corbally}(1994)}]{1994AJ....107..742G}
{Gray}, R.~O., \& {Corbally}, C.~J. 1994, \aj, 107, 742

\bibitem[{{Griffin} \& {Stencel}(2013)}]{2013PASP..125..775G}
{Griffin}, R.~E., \& {Stencel}, R.~E. 2013, \pasp, 125, 775

\bibitem[{{Gyldenkerne}(1970)}]{1970VA.....12..199G}
{Gyldenkerne}, K. 1970, Vistas in Astronomy, 12, 199

\bibitem[{{Han} {et~al.}(2003){Han}, {Podsiadlowski}, {Maxted}, \&
  {Marsh}}]{2003MNRAS.341..669H}
{Han}, Z., {Podsiadlowski}, P., {Maxted}, P.~F.~L., \& {Marsh}, T.~R. 2003,
  \mnras, 341, 669

\bibitem[{{Han} {et~al.}(2002){Han}, {Podsiadlowski}, {Maxted}, {Marsh}, \&
  {Ivanova}}]{2002MNRAS.336..449H}
{Han}, Z., {Podsiadlowski}, P., {Maxted}, P.~F.~L., {Marsh}, T.~R., \&
  {Ivanova}, N. 2002, \mnras, 336, 449

\bibitem[{{Henden} {et~al.}(2016){Henden}, {Templeton}, {Terrell}, {Smith},
  {Levine}, \& {Welch}}]{2016yCat.2336....0H}
{Henden}, A.~A., {Templeton}, M., {Terrell}, D., {et~al.} 2016, VizieR Online
  Data Catalog, 2336

\bibitem[{{Hoard} {et~al.}(2010){Hoard}, {Howell}, \&
  {Stencel}}]{2010ApJ...714..549H}
{Hoard}, D.~W., {Howell}, S.~B., \& {Stencel}, R.~E. 2010, \apj, 714, 549

\bibitem[{{Howell} {et~al.}(2014){Howell}, {Sobeck}, {Haas}, {Still},
  {Barclay}, {Mullally}, {Troeltzsch}, {Aigrain}, {Bryson}, {Caldwell},
  {Chaplin}, {Cochran}, {Huber}, {Marcy}, {Miglio}, {Najita}, {Smith},
  {Twicken}, \& {Fortney}}]{2014PASP..126..398H}
{Howell}, S.~B., {Sobeck}, C., {Haas}, M., {et~al.} 2014, \pasp, 126, 398

\bibitem[{{Huang}(1965)}]{1965ApJ...141..976H}
{Huang}, S.-S. 1965, \apj, 141, 976

\bibitem[{{Ji} {et~al.}(2013){Ji}, {Fisher}, {Garc{\'{\i}}a-Berro},
  {Tzeferacos}, {Jordan}, {Lee}, {Lor{\'e}n-Aguilar}, {Cremer}, \&
  {Behrends}}]{2013ApJ...773..136J}
{Ji}, S., {Fisher}, R.~T., {Garc{\'{\i}}a-Berro}, E., {et~al.} 2013, \apj, 773,
  136

\bibitem[{{Kemp} {et~al.}(1986){Kemp}, {Henson}, {Kraus}, {Beardsley},
  {Carroll}, {Ake}, {Simon}, \& {Collins}}]{1986ApJ...300L..11K}
{Kemp}, J.~C., {Henson}, G.~D., {Kraus}, D.~J., {et~al.} 1986, \apjl, 300, L11

\bibitem[{{Kharchenko} \& {Roeser}(2009)}]{2009yCat.1280....0K}
{Kharchenko}, N.~V., \& {Roeser}, S. 2009, VizieR Online Data Catalog, 1280

\bibitem[{{Kloppenborg} {et~al.}(2010){Kloppenborg}, {Stencel}, {Monnier},
  {Schaefer}, {Zhao}, {Baron}, {McAlister}, {ten Brummelaar}, {Che},
  {Farrington}, {Pedretti}, {Sallave-Goldfinger}, {Sturmann}, {Sturmann},
  {Thureau}, {Turner}, \& {Carroll}}]{2010Natur.464..870K}
{Kloppenborg}, B., {Stencel}, R., {Monnier}, J.~D., {et~al.} 2010, \nat, 464,
  870

\bibitem[{{Kloppenborg} {et~al.}(2015){Kloppenborg}, {Stencel}, {Monnier},
  {Schaefer}, {Baron}, {Tycner}, {Zavala}, {Hutter}, {Zhao}, {Che}, {ten
  Brummelaar}, {Farrington}, {Parks}, {McAlister}, {Sturmann}, {Sturmann},
  {Sallave-Goldfinger}, {Turner}, {Pedretti}, \&
  {Thureau}}]{2015ApJS..220...14K}
{Kloppenborg}, B.~K., {Stencel}, R.~E., {Monnier}, J.~D., {et~al.} 2015, \apjs,
  220, 14

\bibitem[{{Kochanek} {et~al.}(2017){Kochanek}, {Shappee}, {Stanek}, {Holoien},
  {Thompson}, {Prieto}, {Dong}, {Shields}, {Will}, {Britt}, {Perzanowski}, \&
  {Pojma{\'n}ski}}]{2017PASP..129j4502K}
{Kochanek}, C.~S., {Shappee}, B.~J., {Stanek}, K.~Z., {et~al.} 2017, \pasp,
  129, 104502

\bibitem[{{Kochukhov} {et~al.}(2004){Kochukhov}, {Bagnulo}, {Wade}, {Sangalli},
  {Piskunov}, {Landstreet}, {Petit}, \& {Sigut}}]{2004A&A...414..613K}
{Kochukhov}, O., {Bagnulo}, S., {Wade}, G.~A., {et~al.} 2004, \aap, 414, 613

\bibitem[{{Kochukhov} \& {Wade}(2010)}]{2010A&A...513A..13K}
{Kochukhov}, O., \& {Wade}, G.~A. 2010, \aap, 513, A13

\bibitem[{{Kopal}(1971)}]{1971Ap&SS..10..332K}
{Kopal}, Z. 1971, \apss, 10, 332

\bibitem[{{Kov{\'a}cs} {et~al.}(2002){Kov{\'a}cs}, {Zucker}, \&
  {Mazeh}}]{2002A&A...391..369K}
{Kov{\'a}cs}, G., {Zucker}, S., \& {Mazeh}, T. 2002, \aap, 391, 369

\bibitem[{{Kuiper} {et~al.}(1937){Kuiper}, {Struve}, \&
  {Str{\"o}mgren}}]{1937ApJ....86..570K}
{Kuiper}, G.~P., {Struve}, O., \& {Str{\"o}mgren}, B. 1937, \apj, 86, 570

\bibitem[{{Kurucz}(2005)}]{2005MSAIS...8..189K}
{Kurucz}, R.~L. 2005, Memorie della Societa Astronomica Italiana Supplementi,
  8, 189

\bibitem[{{Lambert} \& {Sawyer}(1986)}]{1986PASP...98..389L}
{Lambert}, D.~L., \& {Sawyer}, S.~R. 1986, \pasp, 98, 389

\bibitem[{{Leadbeater} {et~al.}(2012){Leadbeater}, {Buil}, {Garrel},
  {Gorodenski}, {Hansen}, {Schanne}, {Stencel}, \&
  {Stober}}]{2012JAVSO..40..729L}
{Leadbeater}, R., {Buil}, C., {Garrel}, T., {et~al.} 2012, Journal of the
  American Association of Variable Star Observers (JAAVSO), 40, 729

\bibitem[{{Lissauer} {et~al.}(1996){Lissauer}, {Wolk}, {Griffith}, \&
  {Backman}}]{1996ApJ...465..371L}
{Lissauer}, J.~J., {Wolk}, S.~J., {Griffith}, C.~A., \& {Backman}, D.~E. 1996,
  \apj, 465, 371

\bibitem[{{Lucy} \& {Sweeney}(1971)}]{1971AJ.....76..544L}
{Lucy}, L.~B., \& {Sweeney}, M.~A. 1971, \aj, 76, 544

\bibitem[{{Marsh} \& {Horne}(1988)}]{1988MNRAS.235..269M}
{Marsh}, T.~R., \& {Horne}, K. 1988, \mnras, 235, 269

\bibitem[{{Maxted} {et~al.}(2001){Maxted}, {Heber}, {Marsh}, \&
  {North}}]{2001MNRAS.326.1391M}
{Maxted}, P.~F.~L., {Heber}, U., {Marsh}, T.~R., \& {North}, R.~C. 2001,
  \mnras, 326, 1391

\bibitem[{{Merrill} \& {Burwell}(1949)}]{1949ApJ...110..387M}
{Merrill}, P.~W., \& {Burwell}, C.~G. 1949, \apj, 110, 387

\bibitem[{{Mikolajewski} \& {Graczyk}(1999)}]{1999MNRAS.303..521M}
{Mikolajewski}, M., \& {Graczyk}, D. 1999, \mnras, 303, 521

\bibitem[{{Miller} {et~al.}(2007){Miller}, {Budaj}, {Richards}, {Koubsk{\'y}},
  \& {Peters}}]{2007ApJ...656.1075M}
{Miller}, B., {Budaj}, J., {Richards}, M., {Koubsk{\'y}}, P., \& {Peters},
  G.~J. 2007, \apj, 656, 1075

\bibitem[{{Moe} \& {Di Stefano}(2017)}]{2017ApJS..230...15M}
{Moe}, M., \& {Di Stefano}, R. 2017, \apjs, 230, 15

\bibitem[{{Muthumariappan} {et~al.}(2014){Muthumariappan}, {Parthasarathy},
  {Leadbeater}, {Potravnov}, {Appakutty}, \& {Jayakumar}}]{2014MNRAS.445.2884M}
{Muthumariappan}, C., {Parthasarathy}, M., {Leadbeater}, R., {et~al.} 2014,
  \mnras, 445, 2884

\bibitem[{{Nelson} \& {Eggleton}(2001)}]{2001ApJ...552..664N}
{Nelson}, C.~A., \& {Eggleton}, P.~P. 2001, \apj, 552, 664

\bibitem[{{Nelson} {et~al.}(2004){Nelson}, {Dubeau}, \&
  {MacCannell}}]{2004ApJ...616.1124N}
{Nelson}, L.~A., {Dubeau}, E., \& {MacCannell}, K.~A. 2004, \apj, 616, 1124

\bibitem[{{Paczynski}(1977)}]{1977ApJ...216..822P}
{Paczynski}, B. 1977, \apj, 216, 822

\bibitem[{{Palacios} {et~al.}(2010){Palacios}, {Gebran}, {Josselin}, {Martins},
  {Plez}, {Belmas}, \& {L{\`e}bre}}]{2010A&A...516A..13P}
{Palacios}, A., {Gebran}, M., {Josselin}, E., {et~al.} 2010, \aap, 516, A13

\bibitem[{{Pandey} {et~al.}(2006){Pandey}, {Lambert}, {Jeffery}, \&
  {Rao}}]{2006ApJ...638..454P}
{Pandey}, G., {Lambert}, D.~L., {Jeffery}, C.~S., \& {Rao}, N.~K. 2006, \apj,
  638, 454

\bibitem[{{Paxton} {et~al.}(2011){Paxton}, {Bildsten}, {Dotter}, {Herwig},
  {Lesaffre}, \& {Timmes}}]{2011ApJS..192....3P}
{Paxton}, B., {Bildsten}, L., {Dotter}, A., {et~al.} 2011, \apjs, 192, 3

\bibitem[{{Paxton} {et~al.}(2015){Paxton}, {Marchant}, {Schwab}, {Bauer},
  {Bildsten}, {Cantiello}, {Dessart}, {Farmer}, {Hu}, {Langer}, {Townsend},
  {Townsley}, \& {Timmes}}]{2015ApJS..220...15P}
{Paxton}, B., {Marchant}, P., {Schwab}, J., {et~al.} 2015, \apjs, 220, 15

\bibitem[{{Peters}(2001)}]{2001ASSL..264...79P}
{Peters}, G.~J. 2001, in Astrophysics and Space Science Library, Vol. 264, The
  Influence of Binaries on Stellar Population Studies, ed. D.~{Vanbeveren}, 79

\bibitem[{{Pojmanski}(1997)}]{1997AcA....47..467P}
{Pojmanski}, G. 1997, ACTAA, 47, 467

\bibitem[{{Pojma{\'n}ski}(2001)}]{2001ASPC..246...53P}
{Pojma{\'n}ski}, G. 2001, in Astronomical Society of the Pacific Conference
  Series, Vol. 246, IAU Colloq. 183: Small Telescope Astronomy on Global
  Scales, ed. B.~{Paczynski}, W.-P. {Chen}, \& C.~{Lemme}, 53

\bibitem[{{Rappaport} {et~al.}(2016){Rappaport}, {Gary}, {Kaye}, {Vanderburg},
  {Croll}, {Benni}, \& {Foote}}]{2016MNRAS.458.3904R}
{Rappaport}, S., {Gary}, B.~L., {Kaye}, T., {et~al.} 2016, \mnras, 458, 3904

\bibitem[{{Rappaport} {et~al.}(2017{\natexlab{a}}){Rappaport}, {Gary},
  {Vanderburg}, {Xu}, {Pooley}, \& {Mukai}}]{2017arXiv170908195R}
{Rappaport}, S., {Gary}, B.~L., {Vanderburg}, A., {et~al.} 2017{\natexlab{a}},
  ArXiv e-prints, 1709.08195

\bibitem[{{Rappaport} {et~al.}(2015){Rappaport}, {Nelson}, {Levine},
  {Sanchis-Ojeda}, {Gandolfi}, {Nowak}, {Palle}, \&
  {Prsa}}]{2015ApJ...803...82R}
{Rappaport}, S., {Nelson}, L., {Levine}, A., {et~al.} 2015, \apj, 803, 82

\bibitem[{{Rappaport} {et~al.}(2017{\natexlab{b}}){Rappaport}, {Vanderburg},
  {Jacobs}, {LaCourse}, {Jenkins}, {Kraus}, {Rizzuto}, {Latham}, {Bieryla},
  {Lazarevic}, \& {Schmitt}}]{2017arXiv170806069R}
{Rappaport}, S., {Vanderburg}, A., {Jacobs}, T., {et~al.} 2017{\natexlab{b}},
  ArXiv e-prints, 1708.06069

\bibitem[{{Rappaport} {et~al.}(2017{\natexlab{c}}){Rappaport}, {Vanderburg},
  {Nelson}, {Gary}, {Kaye}, {Kalomeni}, {Howell}, {Thorstensen}, {Lachapelle},
  {Lundy}, \& {St-Antoine}}]{2017MNRAS.471..948R}
{Rappaport}, S., {Vanderburg}, A., {Nelson}, L., {et~al.} 2017{\natexlab{c}},
  \mnras, 471, 948

\bibitem[{{Rattenbury} {et~al.}(2015){Rattenbury}, {Wyrzykowski},
  {Kostrzewa-Rutkowska}, {Udalski}, {Koz{\l}owski}, {Szyma{\'n}ski},
  {Pietrzy{\'n}ski}, {Soszy{\'n}ski}, {Poleski}, {Ulaczyk}, {Skowron},
  {Pietrukowicz}, {Mr{\'o}z}, \& {Skowron}}]{2015MNRAS.447L..31R}
{Rattenbury}, N.~J., {Wyrzykowski}, {\L}., {Kostrzewa-Rutkowska}, Z., {et~al.}
  2015, \mnras, 447, L31

\bibitem[{{Renson} \& {Manfroid}(2009)}]{2009A&A...498..961R}
{Renson}, P., \& {Manfroid}, J. 2009, \aap, 498, 961

\bibitem[{{Richards} \& {Albright}(1999)}]{1999ApJS..123..537R}
{Richards}, M.~T., \& {Albright}, G.~E. 1999, \apjs, 123, 537

\bibitem[{{Rodriguez} {et~al.}(2016){Rodriguez}, {Stassun}, {Lund}, {Siverd},
  {Pepper}, {Tang}, {Kafka}, {Gaudi}, {Conroy}, {Beatty}, {Stevens}, {Shappee},
  \& {Kochanek}}]{2016AJ....151..123R}
{Rodriguez}, J.~E., {Stassun}, K.~G., {Lund}, M.~B., {et~al.} 2016, \aj, 151,
  123

\bibitem[{{Royer} {et~al.}(2007){Royer}, {Zorec}, \&
  {G{\'o}mez}}]{2007A&A...463..671R}
{Royer}, F., {Zorec}, J., \& {G{\'o}mez}, A.~E. 2007, \aap, 463, 671

\bibitem[{{Schlegel} {et~al.}(1998){Schlegel}, {Finkbeiner}, \&
  {Davis}}]{1998ApJ...500..525S}
{Schlegel}, D.~J., {Finkbeiner}, D.~P., \& {Davis}, M. 1998, \apj, 500, 525

\bibitem[{{Scott} {et~al.}(2014){Scott}, {Mamajek}, {Pecaut}, {Quillen},
  {Moolekamp}, \& {Bell}}]{2014ApJ...797....6S}
{Scott}, E.~L., {Mamajek}, E.~E., {Pecaut}, M.~J., {et~al.} 2014, \apj, 797, 6

\bibitem[{{Shakura} \& {Sunyaev}(1973)}]{1973A&A....24..337S}
{Shakura}, N.~I., \& {Sunyaev}, R.~A. 1973, \aap, 24, 337

\bibitem[{{Shappee} {et~al.}(2014){Shappee}, {Prieto}, {Grupe}, {Kochanek},
  {Stanek}, {De Rosa}, {Mathur}, {Zu}, {Peterson}, {Pogge}, {Komossa}, {Im},
  {Jencson}, {Holoien}, {Basu}, {Beacom}, {Szczygie{\l}}, {Brimacombe},
  {Adams}, {Campillay}, {Choi}, {Contreras}, {Dietrich}, {Dubberley},
  {Elphick}, {Foale}, {Giustini}, {Gonzalez}, {Hawkins}, {Howell}, {Hsiao},
  {Koss}, {Leighly}, {Morrell}, {Mudd}, {Mullins}, {Nugent}, {Parrent},
  {Phillips}, {Pojmanski}, {Rosing}, {Ross}, {Sand}, {Terndrup}, {Valenti},
  {Walker}, \& {Yoon}}]{2014ApJ...788...48S}
{Shappee}, B.~J., {Prieto}, J.~L., {Grupe}, D., {et~al.} 2014, \apj, 788, 48

\bibitem[{{Smith}(1996)}]{1996Ap&SS.237...77S}
{Smith}, K.~C. 1996, \apss, 237, 77

\bibitem[{{Stefanik} {et~al.}(2010){Stefanik}, {Torres}, {Lovegrove}, {Pera},
  {Latham}, {Zajac}, \& {Mazeh}}]{2010AJ....139.1254S}
{Stefanik}, R.~P., {Torres}, G., {Lovegrove}, J., {et~al.} 2010, \aj, 139, 1254

\bibitem[{{Stellingwerf}(1978)}]{1978ApJ...224..953S}
{Stellingwerf}, R.~F. 1978, \apj, 224, 953

\bibitem[{{Strassmeier} {et~al.}(2014){Strassmeier}, {Weber}, {Granzer},
  {Schanne}, {Bartus}, \& {Ilyin}}]{2014AN....335..904S}
{Strassmeier}, K.~G., {Weber}, M., {Granzer}, T., {et~al.} 2014, Astronomische
  Nachrichten, 335, 904

\bibitem[{{Tauris} \& {van den Heuvel}(2006)}]{2006csxs.book..623T}
{Tauris}, T.~M., \& {van den Heuvel}, E.~P.~J. 2006, {Formation and evolution
  of compact stellar X-ray sources}, ed. W.~H.~G. {Lewin} \& M.~{van der Klis},
  623--665

\bibitem[{{Tull} {et~al.}(1995){Tull}, {MacQueen}, {Sneden}, \&
  {Lambert}}]{1995PASP..107..251T}
{Tull}, R.~G., {MacQueen}, P.~J., {Sneden}, C., \& {Lambert}, D.~L. 1995,
  \pasp, 107, 251

\bibitem[{{van Kerkwijk} {et~al.}(2010){van Kerkwijk}, {Chang}, \&
  {Justham}}]{2010ApJ...722L.157V}
{van Kerkwijk}, M.~H., {Chang}, P., \& {Justham}, S. 2010, \apjl, 722, L157

\bibitem[{{Van Rensbergen} \& {De Greve}(2016)}]{2016A&A...592A.151V}
{Van Rensbergen}, W., \& {De Greve}, J.~P. 2016, \aap, 592, A151

\bibitem[{{Vanderburg} \& {Johnson}(2014)}]{2014PASP..126..948V}
{Vanderburg}, A., \& {Johnson}, J.~A. 2014, \pasp, 126, 948

\bibitem[{{Vogt} {et~al.}(2014){Vogt}, {Radovan}, {Kibrick}, {Butler},
  {Alcott}, {Allen}, {Arriagada}, {Bolte}, {Burt}, {Cabak}, {Chloros},
  {Cowley}, {Deich}, {Dupraw}, {Earthman}, {Epps}, {Faber}, {Fischer}, {Gates},
  {Hilyard}, {Holden}, {Johnston}, {Keiser}, {Kanto}, {Katsuki}, {Laiterman},
  {Lanclos}, {Laughlin}, {Lewis}, {Lockwood}, {Lynam}, {Marcy}, {McLean},
  {Miller}, {Misch}, {Peck}, {Pfister}, {Phillips}, {Rivera}, {Sandford},
  {Saylor}, {Stover}, {Thompson}, {Walp}, {Ward}, {Wareham}, {Wei}, \&
  {Wright}}]{Vogt2014}
{Vogt}, S.~S., {Radovan}, M., {Kibrick}, R., {et~al.} 2014, \pasp, 126, 359

\bibitem[{{Webbink}(1984)}]{1984ApJ...277..355W}
{Webbink}, R.~F. 1984, \apj, 277, 355

\bibitem[{{Wolff}(1981)}]{1981ApJ...244..221W}
{Wolff}, S.~C. 1981, \apj, 244, 221

\bibitem[{{Zola} {et~al.}(1994){Zola}, {Hall}, \&
  {Henry}}]{1994A&A...285..531Z}
{Zola}, S., {Hall}, D.~S., \& {Henry}, G.~W. 1994, \aap, 285, 531

\bibitem[{{Zucker} \& {Mazeh}(1994)}]{1994ApJ...420..806Z}
{Zucker}, S., \& {Mazeh}, T. 1994, \apj, 420, 806

\end{thebibliography}
